\documentclass[aps,pre,twocolumn,showpacs,amssymb]{revtex4}

\usepackage[dvips]{graphicx}
\usepackage{dcolumn}
\usepackage{bm}
\def\be{\begin{equation}}
\def\en{\end{equation}}

\def\gdot{\dot{\gamma}}
\def\p{\partial} 
\newcommand{\av}[1]{\langle{#1}\rangle}
\def\gs{\gtrsim}
\def\ls{\lesssim}
\newcommand{\bi}[1]{\mbox{\boldmath$#1$}}
\def\p{\partial}
\def\bea{\begin{eqnarray}}
\def\ena{\end{eqnarray}}

\usepackage{color}

\begin{document}


\title{Plastic deformations in 
crystal, polycrystal, and glass\\  
in  binary mixtures 
 under shear: Collective yielding}



\affiliation{Department of Physics, Kyoto University, Kyoto 606-8502,
Japan}

\author{Hayato Shiba}
\affiliation{Institute for Solid State Physics, University of Tokyo, Chiba 277-8581, Japan}
\affiliation{Department of Physics, Kyoto University, Kyoto 606-8502,
Japan}

\author{Akira Onuki}
\affiliation{Department of Physics, Kyoto University, Kyoto 606-8502,
Japan}


\date{\today}

\begin{abstract}
Using molecular dynamics simulation, 
we examine the dynamics 
of crystal, polycrystal, and glass 
in  a Lennard-Jones  binary mixture 
composed of small and large particles  in two dimensions. 
The crossovers occur 
among these states  as  the composition $c$ is varied at fixed 
size ratio.  
Shear is applied to a system of 9000 particles 
in contact with moving  boundary layers composed of  1800 particles.  
The particle configurations 
 are visualized with  a sixfold orientation angle $\alpha_j (t)$  and    
 a disorder variable $D_j(t)$ defined for  particle $j$, 
where the latter  represents  the deviation from  hexagonal order. 
Fundamental plastic elements are classified into dislocation gliding 
 and grain boundary sliding.  
At any $c$,  large-scale  yielding events 
occur on the acoustic time scale. Moreover, they  multiply 
occur in  narrow fragile areas, forming shear bands.   
   The  dynamics of plastic flow  
is highly hierarchical with a wide range 
of time scales for slow shearing. 
We also clarify the relationship between the 
shear stress averaged in the bulk region 
and the wall stress applied at the boundaries. 
\end{abstract}



\maketitle


\section{Introduction}

A wide variety    of 
rheological problems 
are well known in crystal, polycrystal, 
and glass, where  plastic deformations  
are induced  by structural changes 
with increasing  applied stress \cite{Haasen,Spaepen_review}.  
In crystal, dislocations play a major role 
in plasticity \cite{Friedel,dislocation,postmortem,KubinD}. 
They appear and grow   around relatively fragile 
objects such as  point defects, preexisting 
dislocations, and grain boundaries. 
Viscoplastic deformations  under applied stress 
then   involve  the motion of a large number of
interacting dislocations.  
Furthermore,  in multi-phase alloys with domain structures 
or precipitates, dislocations can appear at the interfaces  to  grow 
into softer regions \cite{Haasen,Friedel,pollock,Minami}. 
In polycrystal, plastic deformations 
can also be induced by sliding  motions of the particles at 
grain boundaries \cite{Hahn}, as  studied by 
 molecular dynamics (MD) simulations 
\cite{Yip,Sch,Yama,Swy,Jack,HamanakaShear}. 
In crystal and polycrystal,  plastic events take place  as  
bursts or avalanches spanning  wide ranges of space and time 
scales  as observed in  
acoustic emission experiments \cite{dislocation} 
and by transmission electron microscopy \cite{postmortem}. 

Much attention has  also been paid to 
rheology in  structurally disordered systems,  
including supercooled liquids and glass 
\cite{Simmons,yo1,An,Cates,Barrat-b,Ber,Barrat2D,Falk2D,Lama,Lacks,Pro,Fu,Ma},
foam and  microemulsion systems \cite{Okuzono,Durian,Pine}, 
colloid suspensions \cite{Poon,Weiz}, and 
granular materials \cite{granular1,Behringer,granular2,Da,Bonn,ha}. 
We mention an  early experiment on glass-forming fluids by  
 Simmons {\it et al} \cite{Simmons}, who  
 found strong shear-thinning behavior 
of the viscosity  expressed as $\eta(\gdot) 
 \cong \eta (0)/(1+\gdot \tau_{\eta} )$ in  soda-lime-silica glasses 
under shear with rate $\gdot$. 
Here $\tau_\eta$ is a rheological time 
of the order of  the structural relaxation 
time $\tau_\alpha$, where  $\tau_\alpha$ grows strongly 
 from a microscopic to 
macroscopic time as the glass transition is approached 
\cite{Binderbook}. This behavior 
has been reproduced in subsequent MD 
 simulations (mostly in another 
expression $\eta\sim \gdot^{-a}$ with $a \sim 0.8$).  
In molecular glasses   this   nonlinear regime emerges 
for  $\gdot \tau_{\alpha}>1$, where 
 shear  accelerates  the 
rearrangement of particle configurations 
in jammed states \cite{yo1}.  
As  a closely  related problem,   
understanding of mechanical properties of amorphous 
metals at high strains is of great 
 technological importance in metallurgy  
\cite{Acta,Haasen,HSChen,Spaepen,Argon,Egami,Langer,Spaepen_review,Zink}.

Numerous MD simulations 
on quiescent binary particle systems 
\cite{Takeuchi,Hiwatari,Harrowell,yo,Kob,Donati,Glo,Dol,Ha,Biroli,Kawasaki,HamaOnuki} revealed that  the glass dynamics  
is   highly heterogeneous on mesoscopic spatial  scales.
In model  amorphous alloys, 
Takeuchi {\it et al.}\cite{Takeuchi} observed 
mesoscopic   heterogeneity in atomic 
motions in an  applied  strain. 
After early findings     
by  Muranaka and Hiwatari 
\cite{Hiwatari} 
and by  Harrowell and coworkers \cite{Harrowell}, 
Yamamoto and one of 
the present authors \cite{yo}  
examined breakage of appropriately 
defined bonds and identified  relatively active regions. 
The broken  bonds accumulated in long  time intervals 
are heterogeneous such that their 
structure factor may be fitted to 
the Ornstein-Zernike form 
$\propto 1/(1+k^2\xi^2)$, 
where the wave number $k$ 
is smaller than the inverse  
particle size and the  correlation length $\xi$ 
 grows  with 
lowering  the temperature $T$. 
 Glotzer  {\it et al.} \cite{Kob,Donati,Glo} 
pointed out relevance of  stringlike clusters of mobile 
particles whose lengths increase at low temperatures. 
Afterwards, 
some authors have claimed the presence of 
  correlation  between 
 the  structural heterogeneity in the particle configurations 
and  the dynamic heterogeneity on long time scales 
\cite{Ha,HamaOnuki,Kawasaki}. Recently significant heterogeneity 
has  been found in the elastic moduli in glass 
\cite{Yoshimoto,Barrat-small}, which is the origin 
of  nonaffine  elastic displacements  
for very  small strains.

In glass under shear, 
the  dynamic heterogeneity from the bond breakage 
 becomes  short-ranged  as if a sheared 
state  is mapped onto  a quiescent state 
at a higher temperature \cite{yo1,An}. 
However, some MD simulations  on glass 
\cite{Barrat-b,Falk2D,Fu} realized  
organization of    ``shear bands'' with localization of 
velocity gradients \cite{band} 
 on the system-size  scale.  
Such  bands were  along the flow direction  
in Refs.\cite{Barrat-b,Falk2D}, while  
Furukawa {\it et al.}\cite{Fu} 
 realized shear  bands transiently (with finite life times) 
equally in the flow and velocity-gradient 
directions in 2D using the Lees-Edwards  boundary condition. 
In experiments on amorphous solids at low $T$, 
shear bands have been observed under 
uniaxial stress  above a yield stress 
\cite{Acta,Haasen,Spaepen_review,exp-band,Jing}.  
The width of shear bands is microscopic in the initial stage \cite{Jing}
 but can grow into micrometer  sizes.
Shear bands under uniaxial stress 
 were realized in MD simulations \cite{Deng,Falk3D,bandB,Robbins}  and in
simulations of  2D phenomenological  models 
\cite{Bulatov,Onuki-plastic}, where the band lines (or planes in 3D) 
make an angle  of $ \pi/4$ with respect to the 
uniaxial  direction.

In this paper, we will 
present MD  simulations extending those 
in our previous papers\cite{HamaOnuki,HamanakaShear}.
We will examine the crossover  
among crystal, polycrystal, and glass 
with varying the composition $c$   
in a model 2D binary mixture  with shear. 
Here the temperature $T$ and the size ratio of 
the diameters of the two components $\sigma_2/\sigma_1$ 
are fixed. Further detailed discussions on  the crossover 
with varying $c$  
will be given elsewhere \cite{PTP}. 
We shall see that  plastic events 
tend to take place   over wide areas 
in short times also in glass, whereas some phenomenological theories 
were based on the  assumption that plastic events  are  spatially 
localized in glass due to the structural disorder
\cite{Spaepen,Argon,Langer}.  
It is worth noting that some simulations have recently 
been performed on the size distribution  of extended 
plastic events in glass \cite{Pro}.  
One of our purposes in  this paper is  to 
unambiguously  visualize 
the formation and growth of  plastic 
deformations  over a wide range of time scales 
in a sufficiently large system.


The organization of this paper is as follows.
In Sec.II,   our  model  
 and  our simulation method will be explained. 
 In Sec.III, numerical results 
will be presented on 
nonlinear rheology with large stress drops 
and  collective yielding on various time scales. 
We will also clarify the relationship 
between the applied wall stress 
and the shear stress averaged 
within the bulk region.

\section{Model and Simulation Method }

We treat two-dimensional (2D)  binary mixtures 
composed of two atomic species 1 and 2, as in our 
previous papers\cite{HamaOnuki,HamanakaShear}.
The particles   interact  via  
truncated Lennard-Jones (LJ) potentials,  
\begin{equation}
v_{\alpha\beta} (r) = 4\epsilon \left[ \left(\frac{\sigma_{\alpha\beta}}{r}\right)^{12} - \left(\frac{\sigma_{\alpha\beta}}{r}\right)^6\right] -C_{\alpha\beta} \quad
\label{eq:LJP}
\end{equation}
which are characterized by the energy $\epsilon$ 
and the interaction lengths    
$\sigma_{\alpha\beta} = 
(\sigma_\alpha +\sigma_\beta )/2$ $(\alpha,\beta=1,2)$. 
Here $\sigma_1$ and $\sigma_2$ represent 
 the soft-core diameters of the two components and 
their ratio is fixed at  $\sigma_2/\sigma_1=1.4$. 
For $r>r_{\scriptsize{\textrm{cut}}} 
=3.2\sigma_1$, we set $v_{\alpha\beta} =0$
and the constant $C_{\alpha\beta}$ 
ensures the continuity of $v_{\alpha\beta}$
at $r=r_{\scriptsize{\textrm{cut}}}$. 
The  mass ratio is fixed at  $m_1/m_2 = (\sigma_1/\sigma_2)^2$.

We divide the system into three 
regions  (see Fig. \ref{fig:DBB1} below). 
In  the bulk region  $-0.5 L<x,y<0.5L$, 
we initially placed   $N=N_1+N_2=$ {{9000}} 
particles. The  composition of the larger particles is 
defined as 
\be 
c=N_2/(N_1+N_2). 
\en 
The volume   $L^2$ of the bulk region 
 is chosen such that the volume fraction 
of the soft-core regions is fixed at $1$  or\cite{fraction}
\begin{equation}
\phi=  (N_1\sigma_1^2 +N_2\sigma_2^2)/L^2 = 1.
\label{eq:VF1}
\end{equation}
For example, $ L = 97.12$ for  $c = 0.05$ 
and $L = 103.58$ for  $c = 0.2$.  
Thus $L\sim 100\sigma_1$ in our simulation. 
We  apply shear flow  by 
the boundary motions of   two boundary layers, 
which are  expressed by  
$-0.6L<y<-0.5 L$ and $-0.5 L<x<0.5 L$ at the bottom and 
by $0.5L<y<0.6 L$ and $-0.5L<x<0.5L$ at the top.  
In  each layer, 
$N_b= 900$ binary particles
with the same composition and size ratio 
were initially placed. They  are attached to it 
by the spring potential,  
\begin{equation}
u_{j} (\bm{r}-\bm{R}_j) 
= \frac{1}{2} K |\bm{r} -\bm{R}_j|^2, 
\label{eq:SPP1}
\end{equation} 
where   $\bm{R}_j$ are pinning points appropriately 
determined  in the  boundary layers (see below). 
The spring constant 
is set equal to $K= {20\epsilon}{\sigma_1^{-2}} $.
These bound particles  also interact with the neighboring  
bound and unbound particles with the common 
Lennard-Jones potentials in Eq. (\ref{eq:LJP}). 
The total potential energy 
 is thus written as 
\be 
U= \sum_{j,k{\in {\rm all}} }\phi_{jk}+ \sum_{j\in b}u_{j}
\en 
The  $\phi_{jk}=v_{\alpha\beta}(|{\bi r}_j-{\bi r}_k|)$ 
is the pair potential   
between the  particles $j$ and $k$, with $j$ and $k$ 
being  either unbound or bound to the walls. 
The $u_j$ is  the spring potential in Eq.(4) between 
the bound particle $j \in b$ 
and the pinning  point ${\bi R}_j$. 
After application of shear flow at $t=0$ the pinning points 
depend on time as
\begin{equation}
\bm{R}_j (t)= \bm{R}_j(0)\pm \frac{1}{2}L\dot{\gamma}t {\bi e}_x,
\label{eq:FPP1}
\end{equation}
where ${\bi e}_x$ is the unit vector along the $x$ axis.
If its $x$ component $X_j(t)$ 
became  larger than $L/2$ in the upper layer 
(smaller than $-L/2$ in the lower layer), it was  
decreased (increased) by $L$. That is, we assumed 
 the periodic boundary condition 
along the $x$ axis. 
In our simulation the unbound particles 
rarely penetrated   into 
the boundary layers 
deeper than  $\sigma_1$.

We  integrated the  equations of motion 
using  the leapfrog algorithm 
under the periodic boundary condition 
along  the $x$ axis.
The time step of integration is $0.002\tau$ with
\begin{equation}
\tau = \sigma_1 \sqrt{m_1/\epsilon}.
\end{equation}
The unbound particles  obeyed  the Newton equations 
$m_j \ddot{\bm{r}}_j =  
- {\partial U}/{\partial\bm{r}_j}$, where 
 $\ddot{\bi r}_j= d^2{\bi r}_j/dt^2$. 
To subtract the heat produced in shear flow,  
we attached   a Nos\`e-Hoover thermostat \cite{nose,Hoover} 
to each boundary layer.  That is, 
the bound particles $j \in {\cal B}$ 
 are governed by 
\be 
m_j \ddot{\bm{r}}_j =  
- \frac{\partial U}{\partial\bm{r}_j} - 
\zeta_{\cal B}  m_j (\dot{\bi r}_j-{\bi v}_{\cal B}), 
\en 
where   $\dot{\bi r}_j= d{\bi r}_j/dt$,      
 ${\cal B}$ represents  the top or bottom boundary, 
and ${\bi v}_{\cal B}(= \pm (L\gdot/2){\bi e}_x)$ 
is the boundary velocity at the top or bottom. 
The  two thermostat  variables   $\zeta_{\rm bot}(t) $ and 
$\zeta_{\rm top}(t) $ obeyed     
\be
\frac{d}{dt} {\zeta}_{\cal B} =  \frac{1}{\tau_{\rm{NH}}^2}
\bigg[ \frac{1}{N_bk_BT}\sum_{j \in {\cal B} } 
\frac{m_j}{2} |{\dot{\bi  r}_j-{\bi v}_{\cal B}|^2} -1\bigg], 
\en
where  $\tau_{\rm{NH}}$ is the thermostat 
characteristic time. 
We set $\tau_{\rm{NH}}= 0.304\tau$ in this paper. 
Then the local temperature 
(the local particle kinetic energy) 
became nearly homogeneous  in the bulk region 
during plastic flow with shear 
$\gdot=10^{-4}\tau^{-1}$ \cite{temp}. Here  
 the thermal diffusion time was  shorter than 
the inverse shear $1/\gdot$. 

We explain how we prepared 
the initial particle configurations to which a shear flow was applied. 
(i) We first equilibrated the bulk 
and boundary  regions independently 
without their mutual interactions. 
The particles in these  regions interacted 
via the Lennard-Jones potentials   (\ref{eq:LJP}) 
 in a liquid state at $T=2\epsilon/k_B$ in
a time interval of $10^3\tau$ under  the periodic 
boundary condition in  the $x$ and $y$ axes. 
(ii) Then we quenched the  system to 
$T=0.2\epsilon/k_B$ and further equilibrated the system 
 for a time interval of $10^3\tau$. 
After  this low-temperature  equilibration, 
we chose the  particle  positions in the 
boundary layers  as  the initial pinning   points 
${\bi R}_j(0)$ in Eq. (\ref{eq:SPP1}) and 
 introduced the spring potential (\ref{eq:SPP1}) 
of the bound particles, 
the LJ potentials (\ref{eq:LJP}) between 
 the bound and unbound particles, and the 
Nos\`e-Hoover thermostats of the boundary layers.  
(iii) We further waited  for time interval of 
$5\times 10^3\tau$ until we detected 
 no appreciable time evolution in 
various thermodynamic quantities. 
After this second low-temperature equilibration, we 
 applied a  shear flow  by
sliding the pinning points as in Eq.(6).

\section{Simulation  results}

\begin{figure*}
\includegraphics[scale=0.67]{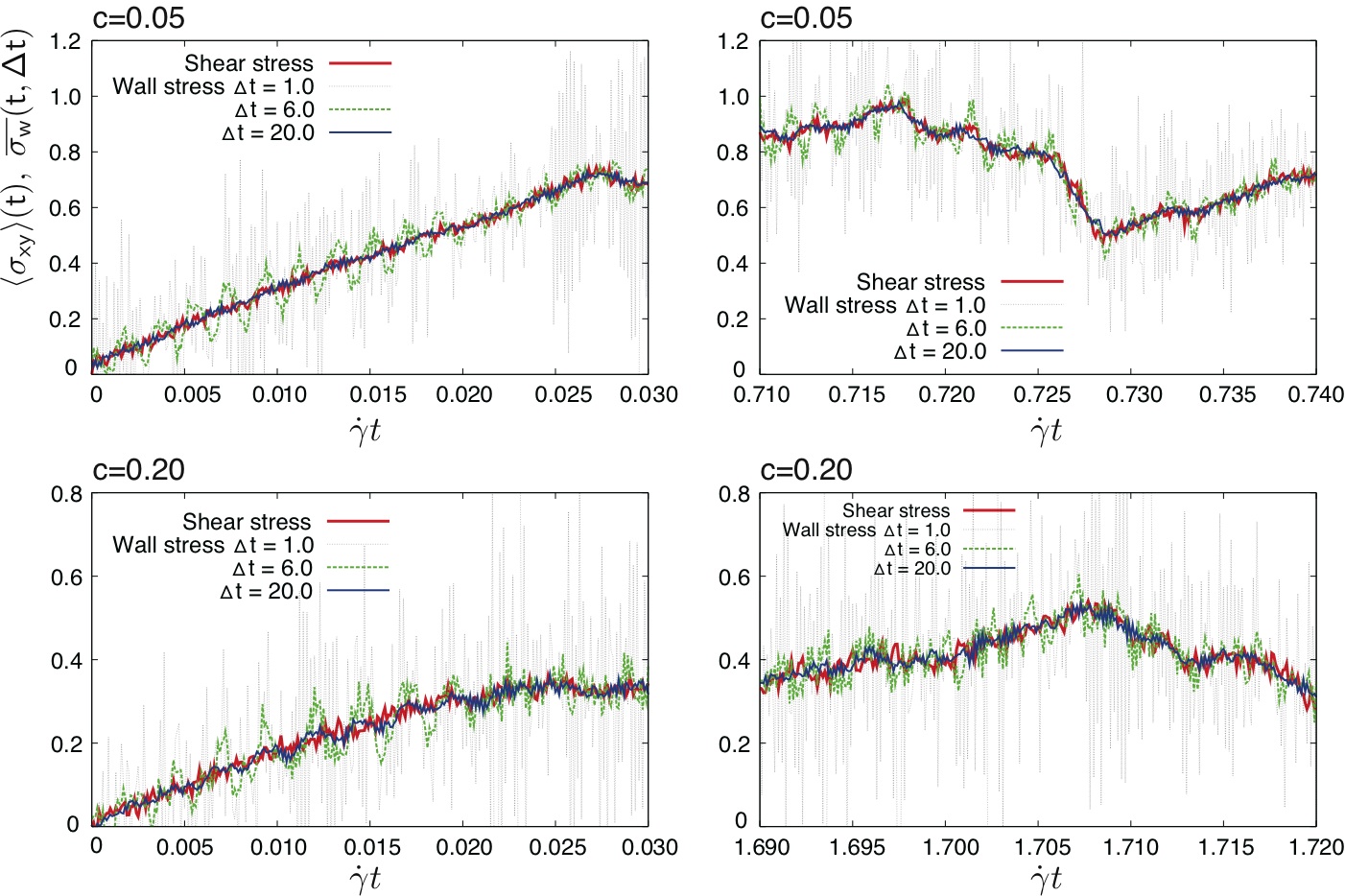}
\caption{(Color online) Smoothed wall stress 
$\bar{\sigma}_{\rm w} (t,\Delta t)$ 
in Eq. (\ref{eq:sigmaw}) and 
average stress $\av{\sigma_{xy}}(t)$ 
under shear $\dot{\gamma} = 10^{-4}$ 
in the initial stage (left) 
and in the plastic flow regime (right), 
where $c= 0.05$ (top) and $c= 0.2$ (bottom) 
with $\Delta t=1,6,$ and 20.  
\label{fig:SCA}
}
\end{figure*}

Hereafter the space coordinates $x$ and $y$, the time $t$,  
the shear rate $\dot\gamma$,  
and the temperature $T$ will be measured 
in units of $\sigma_1$,  $\tau$, $\tau^{-1}$, 
 and $\epsilon /k_B$, respectively, while shear stresses  will 
be in units of $\epsilon\sigma_1^{-2}$.
We will treat  high-density sheared  
 states  at $T=0.2$ and 
 $\dot{\gamma}=10^{-4}$. 
 The   boundary speed ($\sim 0.005$) 
is much slower than the thermal velocity 
($\sim 0.6$). 
There is no tendency of phase separation 
  as in our previous work \cite{HamaOnuki,HamanakaShear}.

\subsection{Previous results }
Jammed particle configurations 
in binary particle systems are 
very complicated depending on $c$ in Eq.(2), $\phi$ in Eq.(3), 
$T$,  and $\sigma_2/\sigma_1$.  
In our previous work \cite{HamaOnuki,PTP}, 
we have examined this problem  
 with varying  $c$ without and with 
shear. We here explain some 
characteristic features in the crossover 
among crystal, polycrystal, and glass 
for  $\sigma_2/\sigma_1=1.4$ and $T=0.2$.

For  small $c$ or $1-c$ less  than a critical 
composition of order  0.05, 
 the overall crystalline order is attained, where 
 the particles of the minority species 
 form localized defects  or 
isolated clusters  with various sizes.   
For $0.05\ls c\ls 0.15$ or  
 $0.8 \ls c\ls 0.95$, 
 polycrystal states are realized, where 
   grain boundaries 
 composed of the two species 
  enclose crystalline domains  
  consisting of the majority species.  
With  increasing $c$ or  $1-c$, 
the grain boudaries are gradually thickened 
into percolated amorphous layers 
enclosing small crystalline domains.  
If $c$ and $1-c$ are not small, glass states 
are eventially realized.

In  polycrystal  between crystal and glass,  
  the grain boundary motions are severely slowed down 
in the presence of  size dispersity $\sigma_2/\sigma_1\neq 1$, 
while the grain boundary motions are rapid 
in one-component systems \cite{2Dmelting}.
As compared to the particles 
within the crystalline regions, 
those  in the grain boundary regions 
are  relatively mobile 
and their collective motions give rise to  
 the dynamic heterogeneity on 
 long time scales \cite{HamaOnuki,Jack}. 
Thus the  origin of the dynamic heterogeneity 
is  rather clear in polycrystal.   
Since small crystalline regions still remain 
 in glass,  the glass dynamics 
 can be understood as the small-grain-size limit of the 
 polycrystal dynamics \cite{HamaOnuki,Jack}. 
   Also varying the degree of disorder,  
Kawasaki {\it et al.} \cite{Kawasaki} 
claimed that  such remaining  
``medium-range crystalline  order'' 
controls both the ease of 
vitrification and nature of the glass transition.

The structural relaxation time $\tau_\alpha$ 
without shear was   of order $10^4$ in glass, 
was  longer in polycrystal,   
and tended  to  infinity in crystal  from the decay of 
the self-time-correlation function \cite{HamaOnuki}. 
The grain boundary motions are much suppressed 
in the presence of size dispersity.

\subsection{Orientation angle $\alpha_j$, 
disorder variable $D_j$, and bond breakage}

We  introduce methods and techniques 
of detecting and visualizing 
structural disorder  and dynamic heterogeneities   
\cite{yo,yo1,HamaOnuki,HamanakaShear}.  
In our 2D systems  a large fraction of the unbound 
particles are enclosed by six particles.  
The local crystalline order may then  be 
represented by a sixfold orientation\cite{NelsonTEXT}.
We define an orientation angle $\alpha_j$ in the range 
$-\pi/6 \le \alpha_j <\pi/6$ 
for each unbound 
particle $j$ using the complex number, 
\bea
\Psi_j &=& 
\sum_{k\in\textrm{\scriptsize{bonded}}} \exp [6i\theta_{jk}]
 \nonumber\\
&=&  |\Psi_j| e^{6i\alpha_j}.
\label{eq:Alpha}
\ena
In the first line,  
the summation is over the particles 
bonded to the particle $j$.
In our case, the two particles $j\in \alpha$ 
and $k\in \beta$ are bonded 
if their distance $r_{jk} =|\bm{r}_j -\bm{r}_k|$ is shorter than 
$1.5\sigma_{\alpha\beta}$.  
The $\theta_{jk}$ is the angle of the relative vector
$\bm{r}_j -\bm{r}_k$ with 
respect to the $x$ axis. The second line is the definition of 
$\alpha_j$. 
It is convenient to 
introduce  another non-negative-definite 
variable
representing the degree of disorder or the deviation from 
 hexagonal order for each particles $j$ by
\bea
D_j &=& 
\sum_{k\in\textrm{\scriptsize{bonded}}}
|e^{6i\alpha_j}-e^{6i\alpha_k}|^2\nonumber\\
&=& 
2 \sum_{k\in\textrm{\scriptsize{bonded}}} [1-\cos 6(\alpha_j -\alpha_k)].
\label{eq:Disodrv}
\ena
For a perfect crystal at low temperature this quantity arises
from thermal vibrations  and is nearly zero,
but for particles around defects  it assumes large values
in the range 15-20.

In jammed states, particle configuration 
changes can be conveniently visualized 
by using  the method of bond breakage. 
That is, for each particle 
 configuration  at a time $t$, 
a pair of particles $i\in\alpha$ and 
$j\in\beta$ is considered to be bonded if
\begin{equation}
r_{ij}(t) = |\bm{r}_i(t)-\bm{r}_j (t)| \le A_1
\sigma_{\alpha\beta},
\end{equation}
where $\sigma_{\alpha\beta}= (\sigma_{\alpha}+ 
\sigma_{\beta})/2$.  We set $A_1=1.2$; then, 
$A_1\sigma_{\alpha\beta}$ 
is slightly larger than the peak distance of the pair correlation 
functions $g_{\alpha\beta}(r)$. 
 After a time interval $\Delta  t$, 
the bond is regarded
to be broken if 
\begin{equation}
r_{ij}(t + \Delta t)\ge A_2\sigma_{\alpha\beta},
\label{eq:BBOD2}
\end{equation}
where we set 
$A_2=1.5$.   
In the following figures, 
bonds broken 
during a time interval $[t_1,t_2]$ 
will be marked by $\times$ 
at the middle point 
$\frac{1}{2}({\bi r}_i(t_2)+{\bi r}_j(t_2))$ 
of the two particle positions 
at the terminal time $t_2$. 
The broken bond number in the system 
in the time interval $[t, t+\Delta t]$ 
will be denoted by  $\Delta N_b(t)$.

\begin{figure*}
\includegraphics[width=0.84\linewidth]{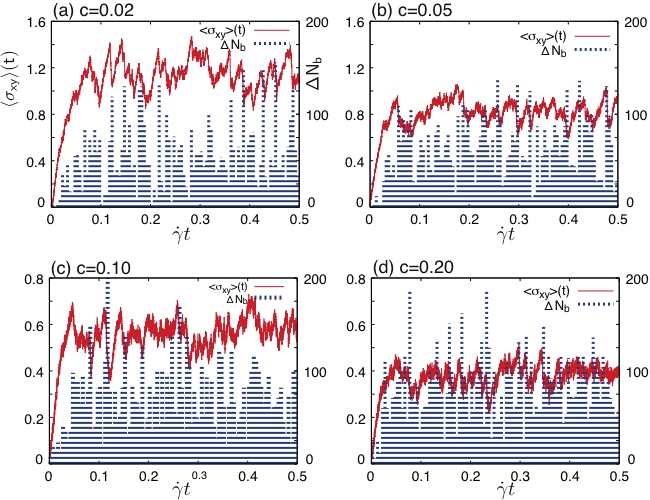}
\caption{(Color online) 
Red solid lines (left scale): 
Average shear stress 
$\av{\sigma_{xy}}(t)$ (in units of $\epsilon\sigma_1^{-2}$) 
vs strain $\dot{\gamma} t$ 
after application of shear  $\dot{\gamma} = 10^{-4}$ 
for (a) $c=0.02$, (b) 0.05, (c) 0.10, and (d) 0.20. 
Blue dotted bars (right scale):  broken bond number $\Delta N_b(t)$ 
in  time interval with width $ 50=5\times 10^{-3}/\dot{\gamma}$ 
in the same runs. Large  stress drops 
are accompanied by  large $\Delta N_b(t)$. 
\label{fig:STS}
}
\end{figure*}

\subsection{Average stress and wall stress}

In the literature of MD simulations of fluids, 
the following  shear stress has been calculated \cite{Onukibook}:   
\be 
\av{\sigma_{xy}}(t)= \frac{-1}{L^2}\bigg[
\sum_j m_j \dot{x}_j\dot{y}_j
-\sum_{jk}\frac{x_{jk}y_{jk}}{2r_{jk}}\phi'_{jk}
\bigg]. 
\label{eq:Stress}
\en 
where we  sum  over 
the unbound particles $j$ and $k$. 
Here $\dot{x}_j=dx_j/dt$ and $\dot{y}_j=dy_j/dt$ 
represent the velocity components, 
 $\phi_{jk}= v_{\alpha\beta}(r_{jk})$ 
($j\in\alpha, k\in \beta$) is    the pair potential  
between $j$ and $k$, and $\phi_{jk}'= \p \phi_{jk}/\p r_{jk}$, 
with   $\bm{r}_j -\bm{r}_k =({x}_j -{x}_k,{y}_j -{y}_k)$ and  
$r_{jk}= |{\bi r}_j-{\bi r}_k|$.  
This quantity is the minus of the 
space average of the $xy$ component 
of the stress tensor $\Pi_{xy}({\bi r},t)$ 
contributed by the unbound particles.

However, it is not clear how $\av{\sigma_{xy}}(t)$ in Eq. (14) 
is related to the experimental shear stress. 
In our geometry, 
$\av{\sigma_{xy}}(t)$  may be related to   
the forces   to  the fluid exerted by the top and bottom 
boundary layers via the springs,  
\bea 
F_{\rm top}(t)&=& K\sum_{j\in {\rm top}} [X_j(t)- x_j], 
\nonumber\\
F_{\rm bot}(t)&=& -K\sum_{j\in {\rm bot}} [x_j- X_j(t)], 
\label{eq:Ftb}
\ena
where the  particle $j$ is bound to 
the top layer in $F_{\rm top}(t)$ and to the bottom  layer 
in $F_{\rm bot}(t) $ with 
 $X_j(t)= X_j(0)\pm \dot{\gamma} t L/2$ 
being  the $x$ component 
of the pinning position ${\bi R}_j(t)$ in Eq. (\ref{eq:FPP1}). 
In shear flow,  the time-dependent 
energy input from the walls to the fluid   is 
given by 
\be 
\dot{W}(t) = L^2 \dot{\gamma} 
\sigma_{\rm w}(t),
\en 
per unit time. The $\sigma_{\rm w}(t) $ is the wall stress written as 
\be 
\sigma_{\rm w}(t) 
= [F_{\rm top}(t)-F_{\rm bot}(t)]/2L.
\label{eq:sigmaw}
\en  
The total particle energy 
$E= \sum_j m_j {\dot{\bi r}}_j^2/2+U$ changes in time as 
\be 
\frac{d }{dt}E = \dot{W}- 
\sum_{\cal B}  \sum_{j\in{\cal B}}  \zeta_{{\cal B}} 
 m_j |{\dot{\bi r}}_j-{\bi v}_{\cal B}|^2,
\en  
where the second  term in the right hand side 
 represents  the energy absorption 
in the boundary layers.

In steady states, 
the time averages  of $\sigma_{\rm w}(t)$  
and $\av{\sigma_{xy}}(t)$ 
should coincide. This  time-averaged value 
$\bar{\sigma}_{\rm w}=\bar{\sigma}_{xy}$  
is equal to the steady-state shear viscosity $\eta(\gdot)$ 
multiplied by $\dot\gamma$. 
However,  $\sigma_{\rm w}(t)$  
consists  of rapid motions of the spring contraction 
and extension   on the time scale of 
\be 
\tau_{\rm sp}= 
\pi (m_1/K)^{1/2} , 
\en  
which is of order $\tau$ in Eq.(7). 
In Fig. \ref{fig:SCA},  
we compare $\av{\sigma_{xy}}(t)$ in Eq. (14)  
and a smoothed wall-stress defined by 
\be 
\bar{\sigma}_{\rm w} (t,\Delta t)
=\frac{1}{\Delta t}\int_{-\Delta t/2}^{\Delta t/2}dt' 
 {\sigma}_{\rm w} (t+t'),
\en  
in the initial  stage and in the plastic flow regime. 
This quantity should coincide with  $\av{\sigma_{xy}}(t)$  
for long-time smoothing in the case  
$\Delta t> \tau_{\rm st}$, where  $\tau_{\rm st}$ 
is a crossover  time of the stress response. 
In our  case, we find  $ \tau_{\rm st}\sim 20\tau$. 
We notice that this $\tau_{\rm st}$  
 is  of the order of the acoustic traversal time, 
\be 
\tau_{\rm ac}=L/c_\perp,
\en 
where   
$c_\perp= (G/{\bar\rho})^{1/2}$ 
is the transverse sound velocity.  
Here  the average mass density $\bar{\rho}
=[m_1(1-c)+m_2c]N/L^2$ 
is of order $ m_1\sigma_1^{-2}$  
and the shear modulus 
$G$  will be  calculated to be of order 
$20 \epsilon\sigma_1^{-2}$ 
in Figs. \ref{fig:STS} and 
\ref{fig:SVE} below, so 
$c_\perp \sim 5\sigma_1/\tau$  and $\tau_{\rm ac}\sim 20\tau$. 
The stress deviations  
arising from local plastic deformations 
propagate 
outwards with sppeds on the order of 
the sound velocity, though  they are 
not small elastic deformations.
In our simulation, 
 we numerically find    the 
following  approximate relation,  
\be 
\sigma_{\rm w}(t) -  
\av{\sigma_{xy}}(t) \cong \frac{1}{L^2}
\frac{d}{dt}\sum_{j\in all}  m_j y_j\dot{x}_j,
\en 
where all the unbound and bound particles 
are summed in  the right hand side. 
See the appendix for more details leading to  Eq. (22).

Furthermore, we consider the consequence of  the energy balance 
Eq. (18) in  steady states, where  
the time average of $dE/dt$ vanishes. 
In the right hand side of Eq.(18), the time average of $\dot W$ 
is equal to $L^2\gdot \bar{\sigma}_{xy}$ in terms of the average stress  
$\bar{\sigma}_{xy}$ (the time average of 
$\av{\sigma_{xy}}(t)$), while   those   
of $\zeta_{\rm top}$ and $\zeta_{\rm bot}$ 
should assume the same value $\bar{\zeta}
=\overline{{\zeta}_{\rm top}}=
\overline{{\zeta}_{\rm top}}$. The kinetic energies  
$\sum_{j\in{\cal B}} 
m_j |{\dot{\bi r}_j}-{\bi v}_{\cal B}|^2/2$ 
of the bound particles fluctuated with 
amplitude of order  $5\%$ of 
 the mean value  $N_bk_BT$ in our simulation. 
Thus, 
\be 
\bar{\zeta}= L^2\gdot \bar{\sigma}_{xy}/4N_b k_BT ,
\en 
where the numerator in the right hand side is the viscous 
heat production.  In  our simulation,   
this balance  was well achieved  
and the temperature was kept  nearly 
homogeneous in the plastic flow regime 
\cite{temp}.  For example, for $c=0.05$, we obtain almost the same 
time averages 
$\overline{{\zeta}_{\rm top}}=0.0010$  and 
 $\overline{{\zeta}_{\rm top}}=0.0011$,  
while the right hand side 
of Eq.(23) is $0.0011$ using  $\bar{\sigma}_{xy}=0.855$ 
in Fig.3.  The thermostat 
variables ${\zeta}_{\rm top}(t)$ and ${\zeta}_{\rm bot}(t)$ 
undergo large temporal fluctuations with 
amplitude of order $ 0.1$  for this case. 
In  equilibrium,  the thermostatic 
variable $\zeta (t)$ (usually attached to all the particles 
under  the periodic boundary condition)   
fluctuates around zero \cite{nose,Hoover}.

\subsection{Nonlinear rheology: plastic  and elastic strains}

In Fig. \ref{fig:STS}, we show  the average shear stress 
$\av{\sigma_{xy}}(t)$ and 
the broken bond numbers $\Delta N_b(t)$ 
as  functions  of 
the strain $\dot{\gamma}t$ after 
application of shear  for 
$c=0.02,  0.05, 0.1$, and $0.2$. 
Here the time average 
and variance of $\av{\sigma_{xy}}(t)$ 
decrease with increasing $c$.
In the initial 
regime $\dot{\gamma}t \ls 0.04$, it increases  linearly  as 
\be 
 \av{\sigma_{xy}}(t) \cong G {\dot{\gamma} t}.
 \label{eq:SModulus}
\en  
This is the definition of 
the shear modulus $G$ in this paper. In glass, it  
should be  considerably smaller  
than the infinite-frequency shear modulus 
 $G_\infty$ introduced by 
Zwanzig and Mountain \cite{Zwanzig}, 
since the  correlation-function 
expression for $G_\infty$  
is based  on  the assumption of affine deformations 
of the particle positions (see the appendix 
of Chapter I of Ref.\cite{Onukibook}). 
However,  in glass, the local elastic moduli are highly 
inhomogeneous and the particle displacements are strongly nonaffine 
even for very small strains \cite{Yoshimoto,Barrat-small}. 
In our simulation $G_\infty/G =2-3$ is obtained in polycrystal and glass. 
The expression for the  transverse sound velocity 
  $c_\perp= (G/\bar{\rho})^{1/2}$  is only approximate  if  use is made of 
$G$ in Eq.(24).

In the subsequent plastic flow  in Fig.2, 
$\av{\sigma_{xy}}(t)$  exhibits 
large temporal fluctuations 
in spite of the fact that 
it is the space average in the bulk region with volume 
$L^2 \sim 10^4$ \cite{Pro,variance}.    The maximum stress drops are 
of  order $0.2$ 
for small $c$ and of order  $0.1$ at $c=0.2$.  
They sometimes take  place  abruptly 
on a time scale of the acoustic 
 time $\tau_{\rm ac}$ 
in Eq.(21). They  are induced  by collective   
configuration changes of  the particle positions 
as will be examined later. 
In Fig.2,  this is  demonstrated from the 
  histograms   of  the broken bonds 
in time interval  with width 
$50=5\times 10^{-3}/\dot{\gamma}$ \cite{yo}. 
In MD  simulations on amorphous systems,  
similar  stress-strain curves have 
been calculated (see Refs.\cite{Barrat2D,Falk2D,Lama}, for example).

In plastic flow, the typical amplitude of the 
stress fluctuations around the mean value 
is inversely proportional to the system length $L$ 
(to the square of the volume in 3D) 
\cite{Lama,Barrat2D,variance}.   In  crystalline systems,  
 the stress-strain curve   
 becomes smooth for macroscopic samples, so that     
the avalanche behavior of the dislocation motions 
was detected  by acoustic emission 
measurements using a piezoelectric transducer 
\cite{dislocation}.  In some 
polycrystalline dilute alloys  
such as Al-4 at.$\%$ Mg, however, 
a noisy stress-strain curve has been observed 
 even for  macroscopic samples 
(the Portevin-Le Chatelier effect)  \cite{Kubin}.

We introduce the plastic and elastic strains by simple 
arguments. In  time interval with width $\Delta t$,  
 the  broken bond number 
$\Delta N_b(t)$ 
and  the plastic strain increment 
$\Delta \gamma_{\rm pl}$ are related  as  
\be 
\Delta N_b \sim N\Delta \gamma_{\rm pl}, 
\en 
where $N$ is the total unbound 
particle number. In the plastic flow regime,  
the average of $\Delta \gamma_{\rm pl}$ 
over many successive time intervals  
should be  equal to the applied 
strain increment $\Delta 
\gamma= \dot{\gamma}\Delta t=5\times 10^{-3}$. 
Thus the time average of $\Delta N_b$ is estimated as 
\be 
\overline{\Delta N_b} 
 \sim N \dot{\gamma} \Delta t, 
\label{eq:NBT1}
\en 
which is consistent with 
Fig. \ref{fig:STS} since  
 the right hand side of Eq. (\ref{eq:NBT1}) 
is 45. The elastic 
strain increment  in each time interval 
is given by  the difference,  
\be 
\Delta \gamma_{\rm el}
= \dot{\gamma}\Delta t- \Delta \gamma_{\rm pl}.
\en  
In Fig. \ref{fig:STS}, when  $\av{\sigma_{xy}}(t)$  
increases  gradually,  $\Delta N_b$   
is relatively small and 
$\Delta \gamma_{\rm el}$ also increases gradually 
up  to of order 
$\dot{\gamma}\Delta t$. On the other hand, 
when  $\av{\sigma_{xy}}(t)$  
drops abruptly,  $\Delta N_b$   
is relatively large and $\Delta \gamma_{\rm el}$ 
is negative. As a result, the time average 
 of the elastic strain increment ${\Delta \gamma_{\rm el}}$
should vanish, since  the time average of the 
elastic strain  ${\gamma_{\rm el}}(t)$  
 remains  a constant $\overline{\gamma}_{\rm el}$  
in plastic flow.  Here we relate  the average 
elastic strain $\overline{\gamma}_{\rm el}$  to the  
 time average of the shear   stress 
${\bar\sigma}_{xy} $ as  
\be 
\overline{\gamma}_{\rm el}\sim 
\bar{\sigma}_{xy} /G.
\label{eq:GMPL}
\en

\begin{figure}
\includegraphics[width=0.80\linewidth]{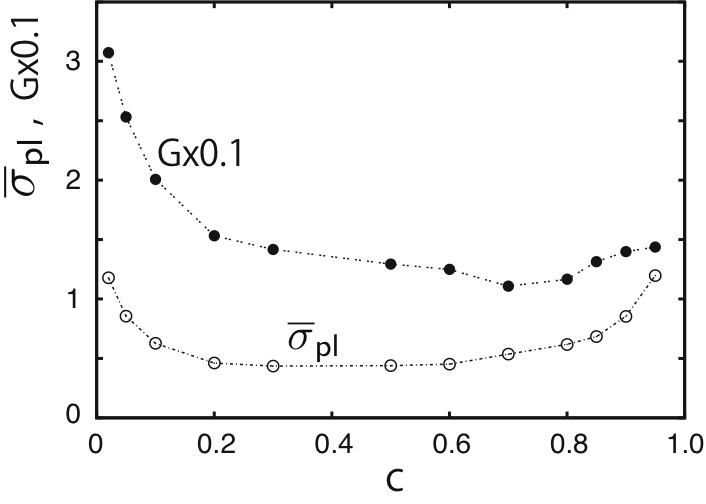}
\caption{Shear modulus $G$ in the initial stage 
in Eq. (\ref{eq:SModulus}) and time-averaged  
stress ${\bar{\sigma}}_{xy}$ in units of $\epsilon\sigma_1^{-2}$ 
 in the plastic flow regime   in Eq. (\ref{eq:PLST}) 
as functions of  $c$   at  $\dot{\gamma} = 10^{-4}$. 
The volume fraction is fixed as in Eq. (\ref{eq:VF1}). 
\label{fig:SVE}
}
\end{figure}

In our case we define the  time-averaged   stress 
${\bar\sigma}_{xy}$ as 
\be 
{\bar{\sigma}}_{xy}    = \frac{1}{t_2-t_1 }\int_{t_1}^{t_2}
dt \av{\sigma_{xy}}(t),
\label{eq:PLST}
\en 
 with $t_1=0.2/ \dot{\gamma}$ and 
$t_2=1/ \dot{\gamma}$ in the plastic flow regime. 
In Fig. \ref{fig:SVE}, we display the shear modulus $G$ 
in the initial stage defined by Eq. (\ref{eq:SModulus}) and 
the  time-averaged   stress 
$\bar{\sigma}_{xy}$ 
in the plastic flow regime. 
For $c\ls 0.5$,  we have 
$G/\bar{\sigma}_{xy} \sim 30$, which then 
leads to a reasonable estimate, 
$\bar{\gamma}_{\rm el}\sim 
0.03$,   from  Eq.(28). 
For $c\gs 0.5$ the ratio gradually 
decreases but remains larger than 10.  
Remarkably, as functions of $c$, both 
$G$  and $\bar{\sigma}_{xy}$ 
 are minimum at intermediate compositions. 
Shikata {\it et al.} \cite{Shikata} 
measured the linear shear viscosity $\eta(c)$ 
of bimodal colloidal suspensions as a function of $c$ 
and found   its  minimum 
at intermediate $c$, where 
the colloid  volume fraction was fixed.

\subsection{Fundamental deformation modes: 
dislocation gliding and grain boundary sliding
}

\begin{figure}
\includegraphics[width=1.0\linewidth]{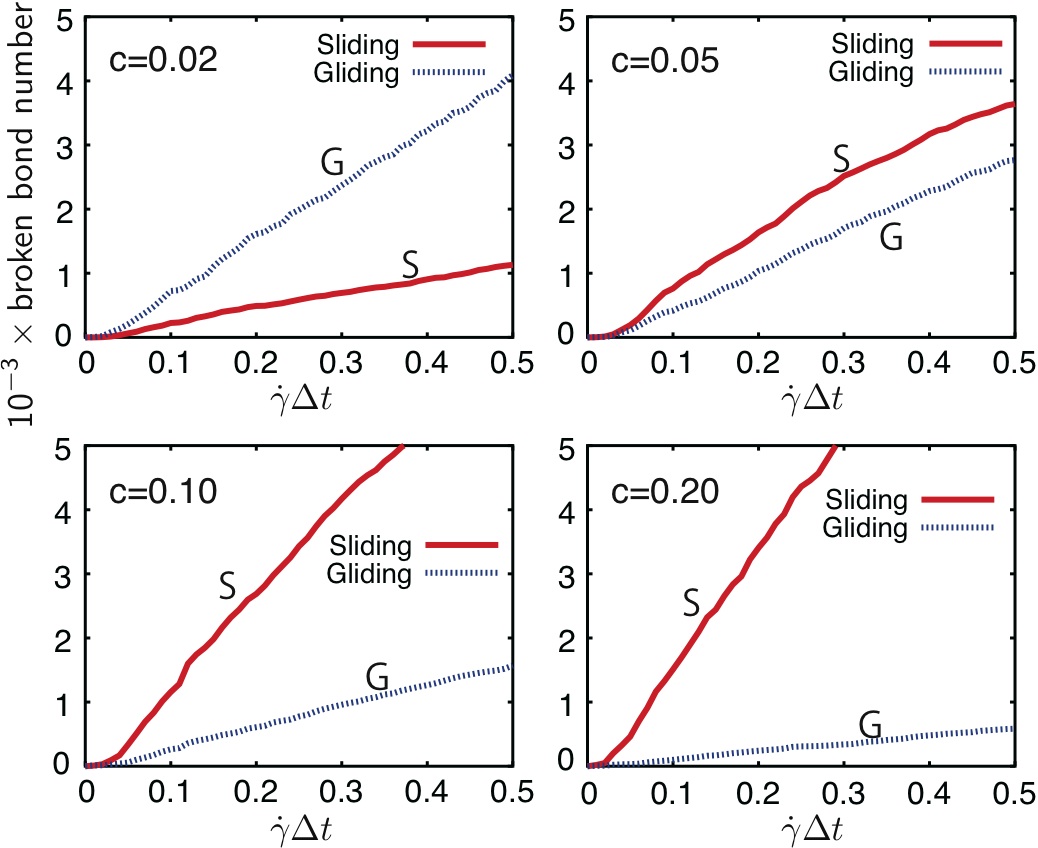}
\caption{(Color online) 
Accumulated numbers  of broken bonds 
in a time interval with width $\Delta t$ 
vs strain increment 
$\dot{\gamma}\Delta t$ for 
$c=0.02, 0.05, 0.10,$ and 0.20. 
Broken bonds  between particles $i$ and $j$  
are classified into 
type G with $D_i+D_j>4$ (red solid lines) 
and type S with $D_i+D_j<4$ 
(blue dotted lines).  Type S broken bonds 
increase with increasing disorder. 
\label{fig:BBODN}
}
\end{figure}

\begin{figure}
\includegraphics[width=0.96\linewidth]{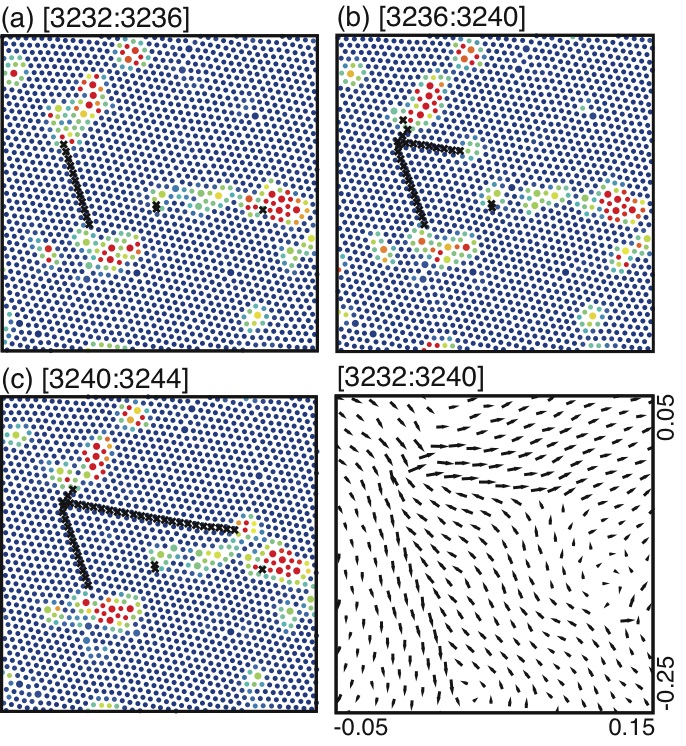}
\caption{(Color online) Successive snapshots of 
broken bonds (${\times}$) generated in the time intervals 
 $[3232, 3236]$ in (a), 
$[3236,3240]$ in (b), 
and $[3240,3244]$ in (c)  
in a crystal state at  $c=0.02$ in the region   
$-0.1L\le x\le 0.2L$ and $0\le y\le 0.3L$. 
Each particle has a color depending 
on its value of the disorder variables $D_j$. 
Right bottom:  Displacement vectors 
${\bi r}_j(t_2)-{\bi r}_j(t_1)$ 
with  $t_1=3232$ and $t_2 = 3240$ 
in the narrower region $-0.05L\le x\le 0.15L$ 
and $0.05 L\le y\le 0.25L$ 
containing  the two slip lines in panel (b). 
An edge dislocation is evident 
at the moving slip end. 
\label{fig:DSlip}
}
\end{figure}

In crystals containing defects and 
polycrystals composed of grains,  
 plastic deformations under applied 
strain  take place in two manners. 
First, dislocations formed around 
defects (grain boundaries, point defects, 
 and preexisting dislocations)  
glide into the grain interior 
with  a speed on the order 
of the  sound velocity  \cite{climb}.  
They eventually form slip planes (lines in 2D) 
bounded by grain boundaries in polycrystals. 
Second,   the particles 
at the grain boundaries intermittently 
undergo sliding motions releasing 
 elastic energies accumulated 
within the crystalline grains. 
In real  3D polycrystals \cite{Hahn,Sch,Yip,Yama,Swy}, 
the dislocation mechanism 
dominates for  grain sizes larger than 
 a critical size $d_c\sim 10$nm, 
while the   sliding mechanism 
dominates for smaller  grain sizes.

In our previous 2D simulation with $N=1000$ 
\cite{HamanakaShear},  we observed 
sliding motions of the particles 
at the grain boundaries in polycrystal, causing  
large stress fluctuations. There, however, 
the sliding motions extended over 
the whole region due to  the small system size.    
In the present  work with  $N=9000$, 
we observe   these  two deformation 
 modes as local events in crystal and polycrystal, 
 while   the sliding motions  
become increasingly short-ranged   
in  glass.

In glass, the local crystalline order
can be  defined only over short distances.
Therefore, fundamental plastic elements have been supposed 
to be  quite   localized in glass 
\cite{Spaepen,Argon,Langer}. 
In a 2D amorphous soap bubble raft, 
Argon and Kuo \cite{Kuo} 
 observed  nucleation of a dislocation pair giving  rise to a
small-scale slip,  though 
 such dislocations did not glide more than
two to three bubble distances.  
With   the size ratio $\sigma_2/\sigma_1$ being  
rather close to unity, Deng {\it et al}. 
\cite{Deng} found extended slips in 2D simulation. 
In glass in our simulation also, short  slips 
may be identified,  where 
 the slip length  does not much 
 exceed the size  of the 
local crystal structure.  
 However, we shall see  (in Figs.12-14  below) that 
such slips successively appear 
in their neighborhood 
to form large-scale  aggregates.

Figure \ref{fig:BBODN}  shows the numbers of 
broken bonds versus 
the strain increment $\dot{\gamma}\Delta t$ 
for $c=0.02, 0.05, 0.1$, and 0.2,  
where $\Delta t$ is  the time interval 
in Eq. (\ref{eq:BBOD2}).  We divide the broken  bonds 
into those with $D_i+D_j>4$ (type S)  
and those with $D_i+D_j<4$ (type G), 
where the bond between 
the  particles   $i$ and $j$ 
is broken. At small $c$,  
the bond breakage 
is mainly caused by the dislocation gliding 
between defects, 
where  the resultant broken bonds are 
type G.   In glass,   
most broken bonds are type S along boundaries of 
small crystalline regions.

Figure \ref{fig:DSlip} displays a typical example of 
dislocation gliding 
along principal crystal axes 
in a crystal state with $c=0.02$. 
In the first panel (left top) 
 a pair of  edge dislocations 
appear around a defect 
and one of them glides into 
the crystal region forming a slip line 
until it is pinned at another defect. 
In the second  panel (right top),
another  edge dislocation is 
 gliding, whose Burgers vector 
is in the slip direction with 
magnitude being equal to 
 the lattice constant $b$ of the hexagonal lattice 
\cite{Friedel}.   
In the third  one (left bottom) 
the second slip is also pinned at a point defect. 
In the right bottom panel,  
the displacements of the particles 
are  around the two slips in the panel (b).

From many such runs  we find   that 
most slips extend nearly  along  the crystal axes 
when they  become  nearly parallel to the 
$x$ or $y$ axis under slow shearing. 
We also notice that there can be two types of slips 
depending on  the directions  
of the Burgers vectors 
at the slip ends.  That is, the particle displacements around a slip  
are either  clockwise  (type C) or 
counterclockwise (type CC). 
In our simulation, slips  
along  the $x$ axis are type C    
and those along  the $y$ are type CC. 
See the right bottom panel of Fig.4 
and the figures in Ref.\cite{Onuki-plastic} 
as examples.

We claim that 
the  preferred directions of the growth  of plastic deformations 
should be  determined by the angle-dependence of the 
elastic energy. 
For example, let us  assume the presence of 
a slip with length $\ell$   under  applied 
shear stress $ \sigma_{xy}^{\rm ext} $.    
We neglect  the crystal structure and the Peierls 
potential \cite{Pei}.    
Using the  Peach-Koehler theory \cite{Peach}, 
we may calculate the   elastic energy of the slip in isotropic 
elasticity as   \cite{Onuki-plastic}, 
\be 
 F_{\rm slip} = \frac{Gb^2}{2\pi} 
 \frac{\ln (\ell/b) }{1-\nu}  
 \mp   \sigma_{xy}^{\rm ext} b \ell \cos(2\varphi) , 
\en 
where $-$ is for  type C,   
 $+$ is for type CC, 
 $b$  is the lattice constant, $G$ is the shear modulus, 
$\nu$ is Poisson's  ratio, and $\varphi$ is the  
angle between the slip direction and  the $x$ axis. 
The first term is the 
elastic  energy of the dislocations, while 
the second term is the work of the applied  force.
 For  simple shear deformation with $ \sigma_{xy}^{\rm ext} >0$,  
 the most favorable slip orientation with the lowest 
$F_{\rm slip} $ is  $\varphi=0$ (along the $x$ axis) 
for type C and $\varphi=\pi/2$ 
(along the $y$ axis) for type CC.  
For such slips,  the negativity 
$F_{\rm slip} <0 $  is eventually realized with increasing 
$\sigma_{xy}^{\rm ext}  \ell$, where the dislocations 
glide  until   they 
 are pinned at  defects. 
In the plastic flow regime,  we should set 
\be 
 \sigma_{xy}^{\rm ext}\sim G \bar{\gamma}_{\rm el} \sim \bar{\sigma}_{xy},
\en  
where  $\bar{\gamma}_{\rm el}$ is introduced in Eq.(28).  
In real crystal, however,   
the critical stress causing
dislocation motion  can strongly depend also on the
orientation of the glide plane  with 
respect to the crystal axes  \cite{Friedel}. 
As stated above,  in our simulation, large-scale 
slip extensions tend to be induced when 
one of the crystal axes become 
nearly parallel to the $x$ or $y$ axis. 
Remarkably, we observed 
the same preferred directions 
in polycrystal and glass (see Fig.6 below). 
Namely, the   grain boundary sliding 
is  easily  induced  along the $x$ or $y$ axis (see Fig.13 also).  

\begin{figure*}
\includegraphics[width=0.80\linewidth]{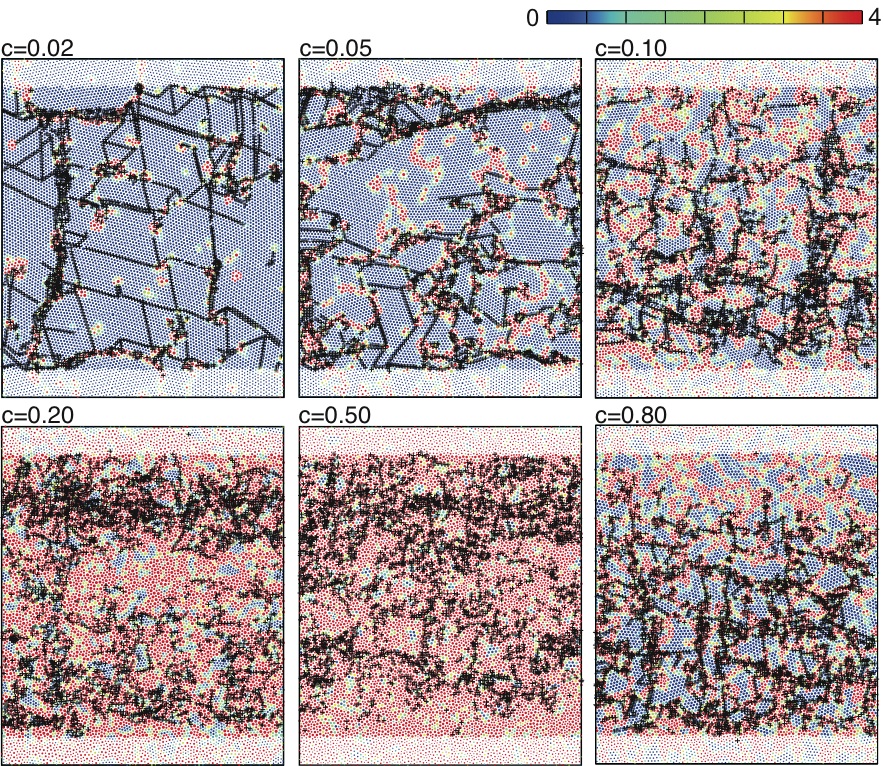}
\caption{
(Color online) 
Broken bonds ($\times$) in the  
time interval  $[8\times 10^3,10^4]$ and 
disorder variable 
 $D_j$  in Eq. (\ref{eq:Disodrv}) at 
the terminal time $t=10^4$  
 of the interval after application of 
 shear $\dot{\gamma} =10^{-4}$, where 
 $c=0.02, 0.05, 0.1, 0.2, 0.5$,  and 0.8 in the six panels. 
Aggregates of broken bonds represent strain localization. 
The colors of the particles are  given 
according to the color bar. 
\label{fig:DBB1}
}
\end{figure*}

Experiments of plastic flow have been  performed under 
 uniaxial stress in metallurgy. The counterpart 
of Eq.(30) is given by 
 \cite{Onuki-plastic}
\be 
 F_{\rm slip} = \frac{Gb^2}{2\pi} 
 \frac{\ln (\ell/b) }{1-\nu}  
 \mp   \sigma_a^{\rm ext}
\sin(2\varphi) ,
\en 
where   $\sigma_a^{\rm ext}
=\av{\sigma_{xx}-\sigma_{yy}}$ is 
the applied uniaxial stress. 
The most favorable direction 
 is given by 
$\varphi=-\pi/4$ for type C and $\varphi=\pi/4$ for type CC. 
These preferred directions have been observed 
in  amorphous metals 
\cite{Acta,exp-band} 
and granular materials \cite{granular1}. 
In simulations, shear bands in these preffered directions  
 have been realized in model  amorphous 
metals and polymers \cite{Deng,Falk3D,bandB,Bulatov,Robbins} 
 and  in a model crystal with weak elastic anisotropy 
 \cite{Onuki-plastic}.

\begin{figure}[t]
\includegraphics[width=0.95\linewidth]{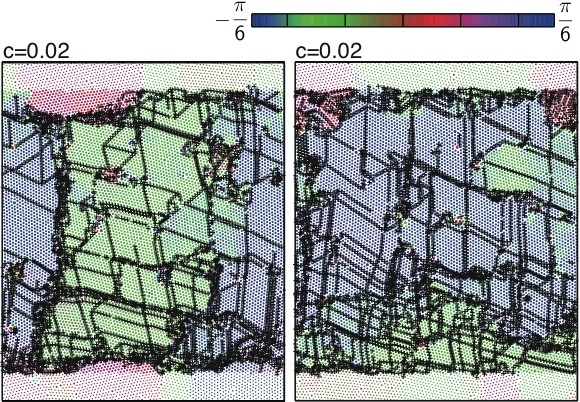}
\caption{(Color online)
Broken bonds ($\times$) in the  
time interval  $[5\times 10^3,10^4]$ and 
orientation angles  $\alpha_j$ in Eq. (\ref{eq:Alpha}) at
the terminal time $t=10^4$ after application of shear for $c=0.02$. 
The colors are  given according to the color bar. 
Left panel is  obtained from the run 
yielding the left top panel in Fig.6. 
Right panel is obtained from 
an independent run, where the strain is localized 
in the lower part.
 \label{fig:ABB1}
 }
\end{figure}

\begin{figure}[t]
\includegraphics[width=1.00\linewidth]{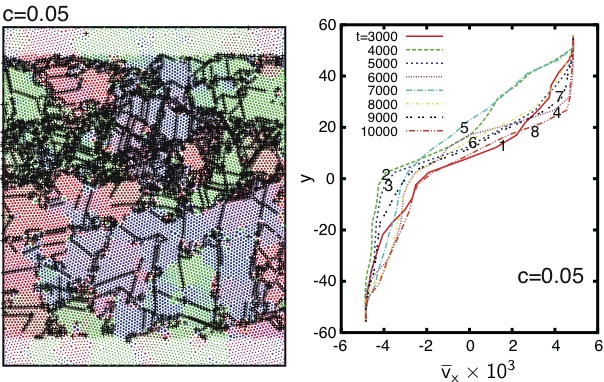}
\caption{ (Color online) 
Left: Broken bonds 
in the  time interval $[5\times 10^3,10^4]$ 
and orientation angles $\alpha_j$ at the terminal time 
$t=10^4$ for $c=0.05$, where 
a long-lived shear band is developed.
Right: Averaged  velocity 
${\bar v}_x(y,t)$ defined by Eqs.(33) and (34)  
at consecutive times 
$t=(2+n)\times 10^3$ ($n=1,\cdots$, and $8$) with $\Delta t=10^3$. 
Strain is  localized in the upper part.
\label{fig:DBBBand}
}
\end{figure}

\begin{figure}
\includegraphics[width=1.00\linewidth]{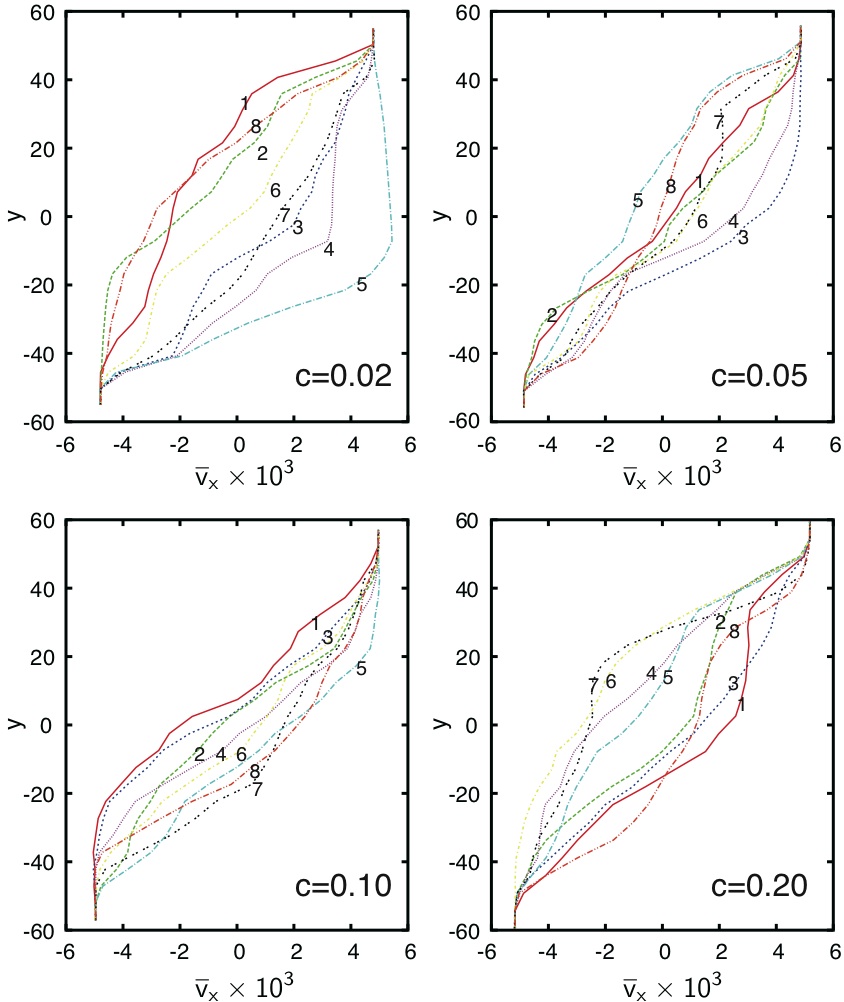}
\caption{(Color online) Consecutive proflies of 
averaged velocity 
$\bar{v}_x(y,t)$ defined by  Eqs. (33) and (34) 
with  $\Delta t = 10^3=0.1/\gdot$  
for $c=0.02, 0.05, 0.10$, and $0.20$. 
Each curve is a  result  from a single run.  
Curves with label $n$ 
are  made  at $t=(2+n)\times 10^3$.  
Space-time fluctuations 
are conspicuous on this time scale of $10^3$ 
at any $c$. 
\label{fig:veloc_prof}
}
\end{figure}
\begin{figure}[h]
\includegraphics[width=0.9\linewidth]{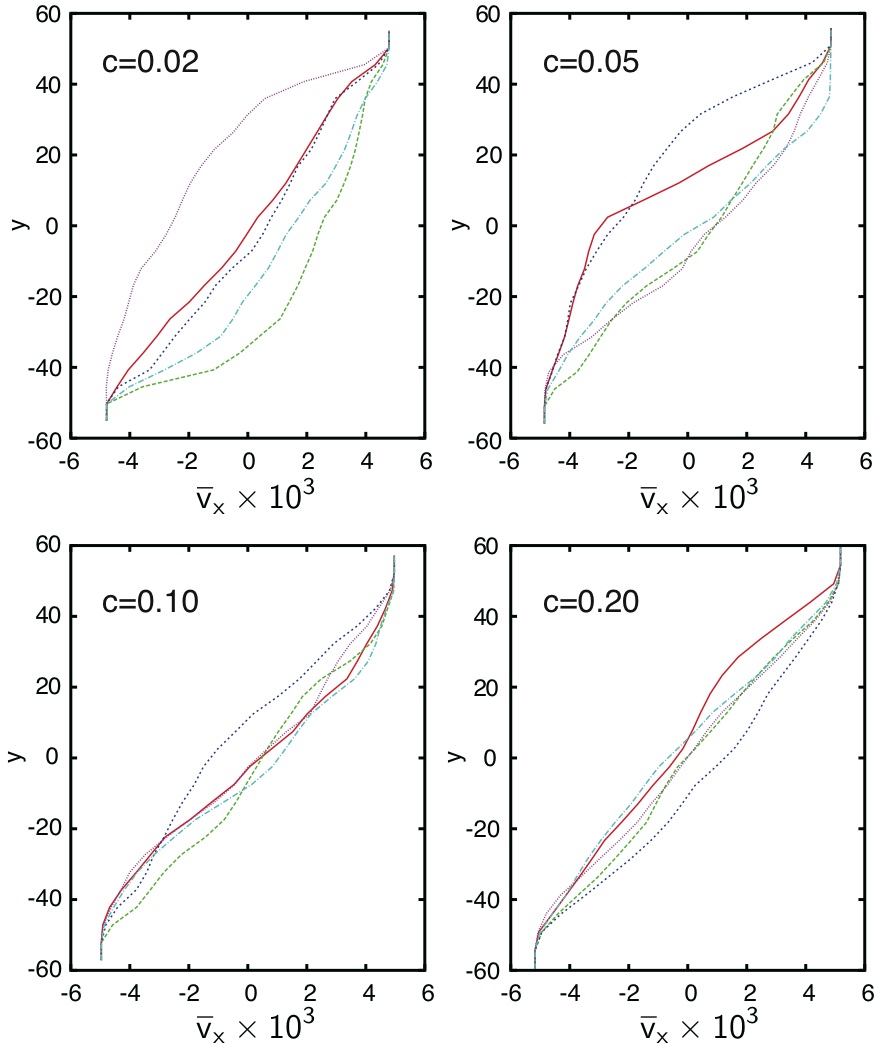}
\caption{(Color online) Averaged  velocity 
 ${\bar v}_x(y, t)$ with longer 
 $\Delta t = 8\times 10^3=0.8/\gdot$ at $t=2\times 10^3$ 
for  $c=0.02, 0.05, 0.1,$ and $0.2$. 
Five curves in each panel are obtained from 
independent runs. For  $c=0.02$ and 0.05, 
they much deviate from 
the linear profile  
depending  on the initial conditions.
For $c=0.1$ and $0.2$, they are nearly 
linear for any  runs.
\label{fig:veloc_prof2} 
}
\end{figure}
\begin{figure*}
\includegraphics[width=0.85\linewidth]{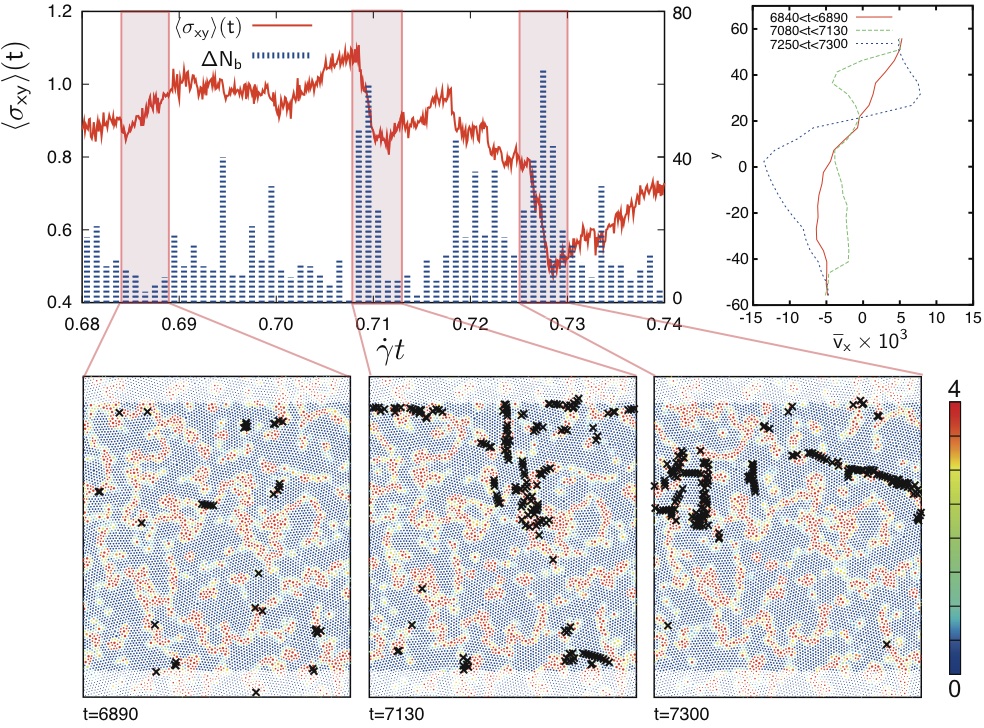}
\caption{(Color online) 
Left top:  
 Shear stress $\langle\sigma_{xy}\rangle(t)$ 
 (red bold line with left scale) 
and broken bond number  $\Delta N_b(t)$ 
in time interval with width 10 
(blue broken bars with right scale) 
as functions of strain $\dot{\gamma} t$  
in polycrystal with 
 $c=0.05$. Bottom: Broken bonds   
at $t=6890, 7130,$ and $ 7300$ in time interval with width 
50 and disordered variable $D_j$ at these times. 
In the first time interval $[6840,6890]$, 
 there is no large-scale plastic deformation, where  
$\langle\sigma_{xy}\rangle(t)$  
linearly grows and  $\Delta N_b(t)$ are small. 
In the second and third  
time intervals $ [7080,7130]$  and $[7250,7300]$, 
$\langle\sigma_{xy}\rangle(t)$  
drops  and  $\Delta N_b(t)$ are large. Right top: 
Averaged  velocity  ${\bar v}_x ( y,t) $  
 at  these times  with  $\Delta t=50$.
\label{fig:SHBB0.05}
}
\end{figure*}

\subsection{Collective yielding 
and averaged velocity on long time scales ($\Delta t \ge 10^3$)}

First, on long time scales with $\Delta t \ge 10^3$, 
we demonstrate that 
plastic deformations  occur collectively 
on large spatial scales,  often 
on  the system-size scale $L$ ($\sim 
100$ here)   at any  $c$, while the type of 
deformations strongly depend on $c$.


Figure \ref{fig:DBB1}  shows  the broken bonds 
in the time interval $[8\times 10^3,10^4]= 
[0.8/\dot{\gamma},1/\dot{\gamma}]$ 
and the disorder variable $D_j$ at the terminal 
time $t=10^4$ for $c=0.02,0.05,0.1,0.2, 0.5$, and 0.8. 
Here a crystal state is realized for 
$c=0.02$, glass  states for 
$c=0.2$ and $ 0.5$, and polycrystal states for the other 
cases.  
At the smallest 
 $c=0.02$,  most bond breakage processes  occur 
 in the form of extended dislocation gliding. 
 A thickened  grain boundary 
in  the left part is produced 
by multiple sliding processes 
(see the left panel of Fig.7 also). 
For $c=0.05$, 
dislocation gliding  occurs with shorter 
 slip lengths. For  $c=0.1$ and 0.8, 
the bond breakage concentrates 
around  the grain boundaries  
resulting in  their sliding, in accord with Fig.5. 
For  $c=0.2$ and 0.5, 
the particle configurations 
are much disordered, but there is  still a tendency of 
the particle motions   along the boundaries of 
 small crystalline regions, as was reported 
in our previous work \cite{HamaOnuki}. 
In all these panels taken for  
 time intervals of $0.2/\gdot$,    heterogeneities of  
broken bonds indicate emergence of large-scale shear bands with 
 high strain localization.  
Furthermore, they mostly  develop  nearly 
 along the $x$ or  $y$ axis. At small $c$, 
 the dislocation mechanism dominates and   these   
features  can be explained by  the slip energy 
in Eq.(30). Remarkably, this orientation  preference 
can be seen also in glass. 
With the boundary layers, the 
horizontal  shear bands  are more frequent than 
the vertical ones in our case, 
while   using the Lees-Edwards  boundary condition  
equally produced    horizontal and 
vertical large-scale shear bands  \cite{Fu}. 
In an experiment on a Laponite suspension \cite{Ma},  
yielding and the velocity field 
sensitively  depended  
on whether the boundary wall is rough or smooth.

Figure  \ref{fig:ABB1}  shows  the broken bonds 
for $c=0.02$ in the longer time interval 
$[5\times 10^3,10^4]= 
[0.5/\dot{\gamma}, 1/\dot{\gamma} ]$ and 
 the  orientation angles  
$\alpha_j$ defined in Eq. (\ref{eq:Alpha}).
The left panel is the result from 
the run producing the left top panel of  Fig.6, 
where the system is divided into 
 two 
  grains since 
the left and right regions are  
connected in our simulation. 
The difference in the 
orientation angles of the two grains 
is about  $\pi/12$.  If we 
compare the left panel   with the left top panel of Fig.6, we recognize 
 accumulation  of dislocation 
gliding and grain boundary 
sliding in  the same  narrow regions.  
The right panel of Fig.7 is 
produced by an independent run, where 
we can see a thickened 
shear band composed of many broken bonds in the 
lower part.

In our system, shear bands  
 disappear on long time scales as in Ref.\cite{Fu}, 
but they are sometimes observed to be long-lived. 
The left panel of Fig.\ref{fig:DBBBand} shows 
such a transient shear band  in the middle region for $c=0.05$,  
which   was  existent  in the time region 
$2\times 10^3 \ls t \ls 1.6\times 10^4$ but   
disappeared  subsequently  on a time scale of $10^3$.  
The right  panel of Fig.\ref{fig:DBBBand} 
gives profiles of an averaged velocity    
$\bar{v}_x(y,t)$ at different times 
in a single run. It  exhibits 
large  gradients along the $y$ axis  in the band 
region with rather small  temporal fluctuations.

To define $\bar{v}_x(y,t)$ in Fig. 8,  
we first integrate  
the $x$ component of the momentum density  
${J}_x(x,y,t)$ 
 in the flow  direction to obtain  
\bea 
{\bar J}_x(y,t)&=& \frac{1}{L} 
\int_{-{L}/{2}}^{{L}/{2}} 
 dx J_x(x,y,t)\nonumber\\
&=&  \frac{1}{L}  \sum_j m_j {\dot x}_j(t) 
 \delta (y-y_j(t)).
\ena  
We  next  average ${\bar J}_x(y,t)$ 
over  space and time intervals with widths $\Delta y$ 
and $\Delta t$ to obtain 
\begin{equation}
{\bar v}_x(y, t) = 
\frac{1}{{\bar \rho}\Delta y\Delta t}
\int^{y+{\Delta y}/{2}}_{{y}-{\Delta y}/{2}}
\hspace{-1mm} {dy'} 
\int^{t+\Delta t}_{t} \hspace{-2mm} {dt'} 
\bar{J}_x(y',t'),
\label{eq:veloc_prof}
\en 
where ${\bar \rho}$ 
is the average mass density. We fix $\Delta y$ at $L/20$. 
However,  we set $\Delta t=10^3=0.1/\gdot$ 
in Figs. 8 and 9,  $\Delta t=8\times 10^3$ in Fig.10, 
and $\Delta t=50$ in Figs.11 and 12 to analyze the 
 hierarchical dynamics of plastic flow 
 spanning   these time scales.

Figure \ref{fig:veloc_prof}  shows eight
consecutive profiles of the  
averaged velocity ${\bar v}_x(y, t)$ 
with  $\Delta t = 10^3=0.1/\gdot$ for 
$c=0.02, 0.05, 0.1,$ and $0.2$ 
in the four panels.  
For these runs, ${\bar v}_x(y, t)$ mostly 
deviates from the linear profile $\gdot y$. 
Their space-time fluctuations  are 
most enhanced  at the smallest composition $c=0.02$. 
Visualized in these time-evolutions 
are emergence and  movement of
fragile areas  on the time scale of $\Delta t= 10^3$.

Figure \ref{fig:veloc_prof2}  
 shows  profiles of the averaged velocity 
${\bar v}_x (y,t)$ with longer $\Delta t= 8\times 10^3= 
0.8/\gdot$ at  $t= 2\times 10^3$ 
 for $c=0.02, 0.05, 0.1,$ and $0.2$. 
In each panel we write five curves obtained from 
independent runs. At $c=0.02$ and 0.05, 
there still remain large deviations from the linear profile,   
sensitively dependent on the  initial conditions 
of simulation.   The curve in red solid line at $c=0.05$ 
is obtained from the run producing the shear band in 
Fig. \ref{fig:DBBBand}. 
For such  low compositions, 
the fragile areas are much extended 
and ${\bar v}_x (y,t)$ changes 
even  on time scales 
longer than $1/\gdot$.  
For higher compositions at $c=0.1$ and 0.2, 
the width $\Delta t
=0.8/\gdot$ is sufficient to yield profiles   
close to the linear one.  That is, with increasing $c$, 
large-scale plastic deformations 
become spatially uncorrelated if they are 
observed on time scales longer 
 than $10^4=1/\gdot$.

\begin{figure*}
\includegraphics[width=0.85\linewidth]{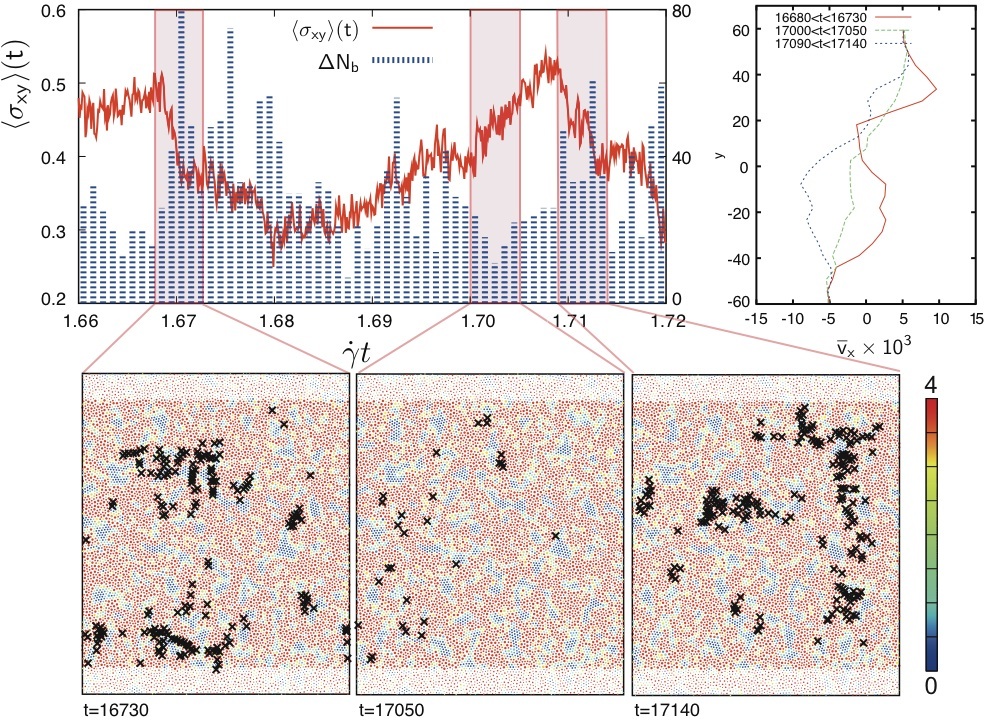}
\caption{(Color online) 
Top left:
Shear stress $\langle\sigma_{xy}\rangle(t)$ 
(red solid line  with  left scale) and  
 broken bond number $\Delta N_b(t)$ 
in time interval with width 10 
(blue broken bars with right scales) as functions of strain  $\dot{\gamma} t$  
in  glass with  $c=0.2$.
Bottom: Broken bonds   
at $t=16730,  17050,$ and $ 17140$ in time interval  with width 
50 and disordered variable $D_j$ at these times. 
Large-scale plastic deformations do not occur 
in the second time interval $[17000,17050]$  
but occur  in the first  and third 
time intervals $ [16680,16730]$  and $[17090,17140]$. 
Right top: Averaged  velocity 
 ${\bar v}_x ( y,t) $  
at  these times with  $\Delta t=50$.
\label{fig:SHBB0.20}
 }
\end{figure*}

\begin{figure}
\includegraphics[width=	1.00\linewidth]{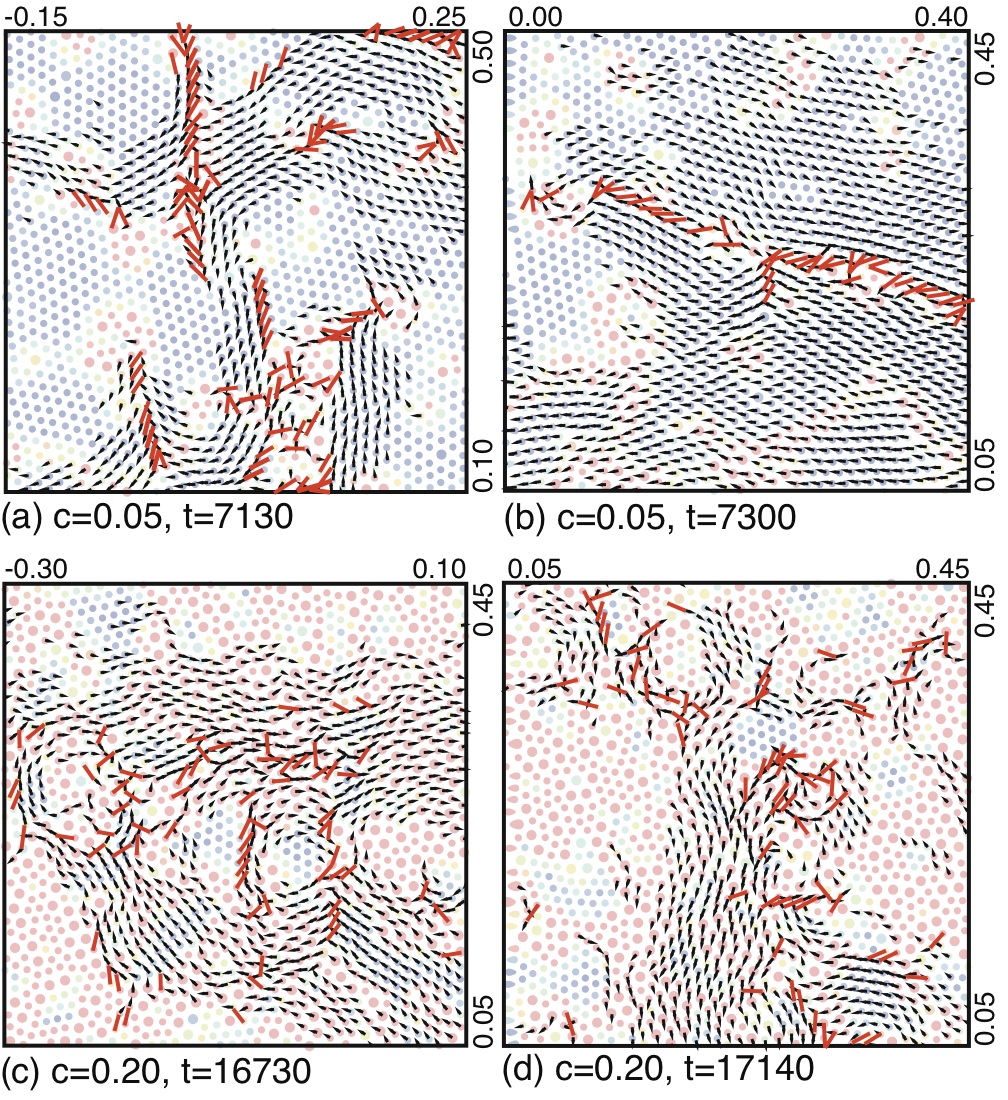}
\caption{(Color online)  Broken bonds (red solid lines) 
and particle displacements with $|{\bi r}_j(t)-{\bi r}_j(t-50)|
>0.3\sigma_1$ (arrows) in 
a time interval $[t-50, t ]$ 
obtained from the runs producing Figs.11 and 12, 
where  (a) $c=0.05$ and 
 $t = 7130$, (b) $c=0.05$ and   $t = 7300$, 
 (c) $c=0.2$ and $t = 16730$, and  
  (d) $c=0.2$ and  $t = 17140$.
  The arrow are from the initial 
  to final positions.  
In  (a) and (b) broken bonds are along 
the grain boundaries. For (c) and (d) time-evolution of 
broken bonds is given in Fig.14.   
Parts of   $16\%$ of the total bulk region are displayed. 
\label{fig:DispV0.20}
}
\end{figure}

\begin{figure}
\includegraphics[width=	0.96\linewidth]{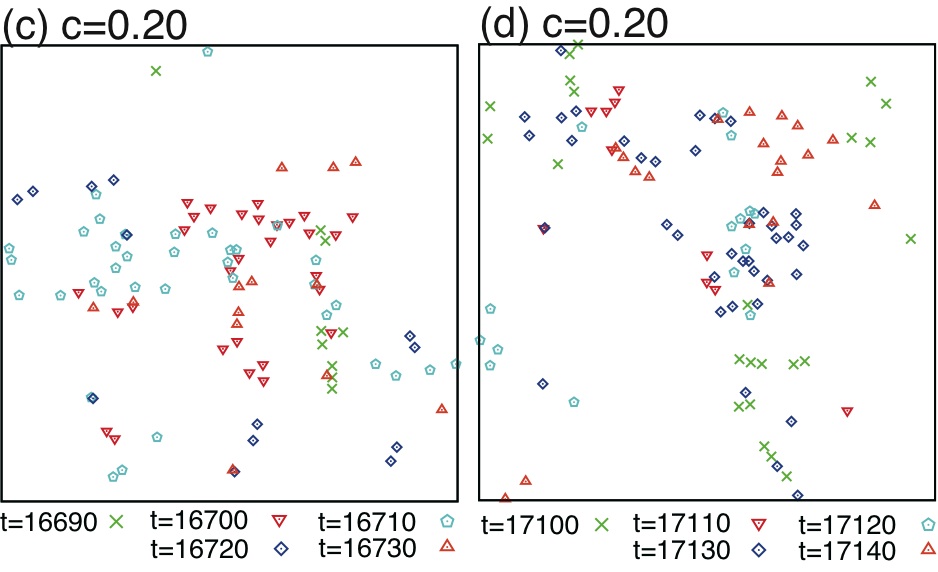}
\caption{(Color online) 
Broken bonds  in   five consecutive time intervals 
$[t-10,t]$ with width $\Delta t=10$ for   $c=0.2$, 
where  $t=16730-10n$ (left) 
 and $t= 17140-10n$ (right)  
 with $0 \le n \le 4$, 
    corresponding to  
 the lower panels (c) and (d) 
 of  Fig.13.    
}
\end{figure}

\begin{figure}
\includegraphics[width=	1.00\linewidth]{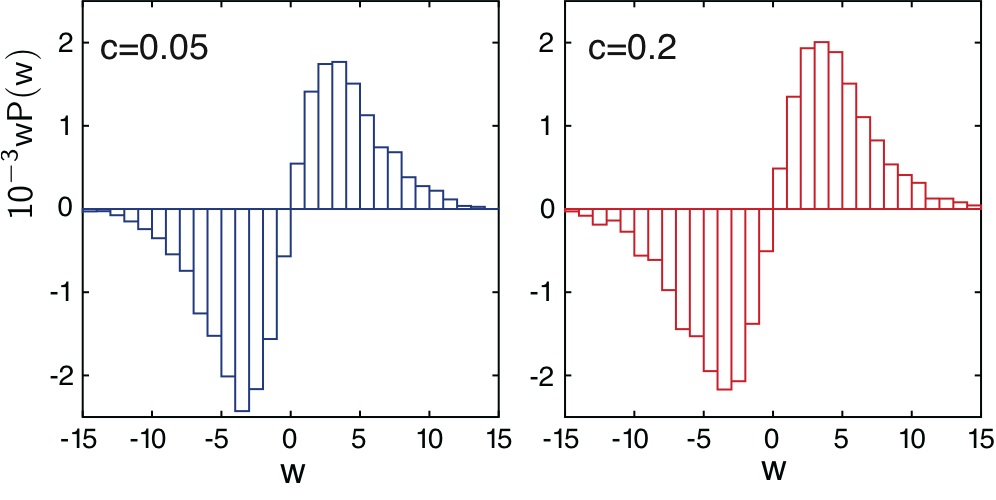}
\caption{(Color online)  Histograms of 
$w P(w)$ defined by Eqs.(37) and (38) 
with $w$ in units of $\epsilon$ in 
the time interval $[7250,7280]$ for $c=0.05$ in Fig.11 (left) 
and the time interval $[17090,17140]$  for $c=0.2$ in Fig.12 (right). 
}
\end{figure}

\subsection{Large stress drops  and collective plastic  events 
on short time scales ($\Delta t \le 50$)}

Figure  2 has shown  that large stress drops  
can occur on a rapid time scale of $\tau_{\rm ac}$ in Eq.(21).  
It is  of great interest how 
 collective plastic deformations develop   
over  large areas  in  short times, 
 involving many particles.

We  investigate this aspect 
for a polycrystal state with 
$c=0.05$ in Fig.\ref{fig:SHBB0.05} 
and for a glass state with 
$c=0.2$ in Fig.12. 
In the upper panels of Figs.11 and 12, 
we present  the shear stress $\langle\sigma_{xy}\rangle(t)$ 
in Eq.(14) and the histograms  of the 
broken bond number $\Delta N_b(t)$ taken 
in  time interval $[t-10,t]$ with width 10.  
The lower  panels  
display snapshots of the broken bonds and the disorder variable $D_j$   
at $t=6890, 7130,$ and $ 7300$ in 
Fig.\ref{fig:SHBB0.05}  
and at $t=16730,  17050,$ and $ 17140$ in Fig.12. 
No large-scale yielding occurs 
in the first time interval $[6840,6890]$ 
in Fig.11 and  in the second time interval  
$[17000,17050]$ in Fig.12, 
where  $\langle\sigma_{xy}\rangle(t)$  
linearly grows and  $\Delta N_b(t)$ remains  small.  
In the other  time intervals,   
correlated bond-breakge events are visualized 
in the forms  of chains in Fig.11 
and aggregates in Fig.12.  They  take place within 
the  interval width 50,  inducing    
  large  $\Delta N_b(t)$ and 
abrupt drops of $\langle\sigma_{xy}\rangle(t)$.
As in Figs.6-8, the large-scale plastic 
deformations tend to 
extend in directions  nearly parallel to the 
$x$ or $y$ axis.  These plastically deformed regions  
are inceptions of shear bands  
and are much longer than the grains in 
polycrystal and the  small remaining 
crystalline regions in glass.

The averaged  velocity 
 ${\bar v}_x ( y,t) $ defined in Eqs. (33) and (34) 
are also calculated   in Figs.11 and 12, where 
the smoothing time interval $\Delta t=50$ 
is much shorter than in Figs.9 and 10.   
For $c=0.05$ in Fig.11,  ${\bar v}_x ( y,t) $ largely  deviates from 
the linear profile even without  collective yielding at $t=6890$. 
This suggests that  short-time deformations can be  
nonaffine  even with weak bond breakage, which is 
in accord with the simulations  in 
Ref.\cite{Yoshimoto,Barrat-small}.  In the third time interval, 
the upper part contains large-scale plastic  deformations and 
move  in the $x$ direction ahead of the upper boundary, while 
the lower part moves back in the negative $x$ direction. 
Also for $c=0.2$ in Fig.12,  ${\bar v}_x ( y,t) $ 
much deviates from the 
linear profile in the presence of collective yielding, 
while it happens to be 
rather close to the linear profile 
in the second interval without  collective yielding.

 In Fig. 13, 
we display the particle displacement,   
\be 
\Delta{\bi r}_j(t)= {\bi r}_j(t)-
{\bi r}_j(t-\Delta t)
\en 
with $\Delta t=50$ around chains or aggregates of the broken bonds, expanding 
active regions of yielding with area $L^2/16$ 
in Figs.11 and 12.  We pick up the particles 
with displacement $|\Delta{\bi r}_j(t)|$ 
longer than $0.3\sigma_1$ 
 suppressing those with  smaller  displacements. 
The  number  of such ``mobile'' particles is 
much larger than the number of 
the broken bonds in the same time interval. 
In Table 1,  
the broken bond number is  about 200 
(about 30 in the narrow regions in  Fig.13), 
while the number of the mobile particles 
is from a few to several thousands. 
The differences of these numbers are  because 
the bond breakage picks up 
the  particle motions most significantly  contributing to 
plastic deformations.  
The broken bonds form long chains for $c=0.05$ and aggregates 
of strings for $c=0.2$  along the $y$ axis 
(left panels) and along the $x$ axis (right panels). 
The displacement vectors   tend to be 
clockwise around horizontal  chains or aggregates 
of the broken bonds 
and counterclockwise 
around vertical  ones. 
In Fig.14, by setting $\Delta t=10$, 
we furthermore show that short 
slip lines in glass successively appear  in their neigborhood 
 for  $c=0.2$.

We then analyze how many particles are involved 
at large stress drops. To this end, we 
rewrite the shear strss in Eq.(14) in  
the sum $\av{\sigma_{xy}}(t)= \sum_j {\cal S}_j(t)/L^2$ with 
\be 
{\cal S}_j(t) = m_j \dot{x}_j\dot{y}_j
-\sum_{k}\frac{x_{jk}y_{jk}}{2r_{jk}}
\phi'_{jk}.
\en 
Since  the stress drop in the time 
interval $[t-\Delta t,t]$ 
is written as $
\sum_j [ {\cal S}_j(t)-{\cal S}_j(t-\Delta t)]/L^2$, 
we may introduce  the particle number 
distribution defined by 
\be 
P(w)=  \sum_j \delta ({\cal S}_j(t)-{\cal S}_j(t-\Delta t)-w).
\en
Then $\int dw  P(w)=N$ and 
\be 
\int dw w P(w)= L^2\bigg[  \av{\sigma_{xy}}(t)- 
\av{\sigma_{xy}}(t-\Delta t)\bigg] 
\en
is equal to the stress drop multiplied by $L^2$. 
In Fig.15, we give histograms of 
$wP(w)$ in regions $n-1/2 <w/\epsilon
<n+1/2$ ($n=0, \pm 1, \pm 2,\cdots$)  
in the time interval $[7250,7280]$ 
for $c= 0.05$ and in the time interval $[17090,17140]$ 
for $c= 0.2$, where the shear stress 
largely drops in Figs.11 and 12. 
Broad distrubutions  of $w$ 
are   due to the 
thermal fluctuations at  high pressure, but 
asymmetry between the negative and positive regions of $w$ 
can be seen  around $w\sim \pm 5\epsilon$. 
As a result,  the integral 
$\int dw w P(w)$ is equal to $ -3.15 \times 10^3$  for $c= 0.05$   
and to $-1.19 \times 10^3$ for $c= 0.2$. 

\begin{table}
\caption{Numbers of broken bonds and mobile 
(largely displaced)  particles 
in the total  bulk region 
at four time intervals with width 
50 in Fig.13. Stress drop 
$\av{\bar{\sigma}_{xy}}(t)- \av{\bar{\sigma}_{xy}}(t-50)$ 
(bottom) is large. }
\label{tab:2d}
\begin{tabular}{ccccccc}
  && (a) & (b) & (c) & (d)  \\
\hline
$c$  && 0.05 & 0.05 & 0.2 & 0.2  \\
\hline
broken bonds  && 170 & 180 & 180 & 200  \\
\hline 
mobile  particles
&& 1820& 5200 & 3500 & 3100 \\
\hline
stress drop 
&& -0.168& -0.278 & -0.100 & -0.133 \\
\hline
\end{tabular}
\end{table}

\section{Summary and remarks }

Though  in two dimensions, we 
 have treated  crystal, polycrystal, and glass  
in a unified manner 
by varying the composition $c$ with  $\sigma_2/\sigma_1=1.4$ 
and $T=0.2\epsilon/k_B$ held fixed. 
Shear flow with $\gdot=10^{-4}$ has been realized with 
 Nos\`e-Hoover thermostats 
attached  to the top and bottom boundary layers.  
We summarize  our main results.\\
(i)  In Sec.IIIC,  we have calculated the  spatially averaged  stress 
$\av{\sigma_{xy}}(t)$ in Eq.(14) and the temporally 
smoothed wall stress $\bar{\sigma}_w(t,\Delta t )$ in Eq.(20) as 
in Fig.1, which  tend to coincide with increasing the 
smoothing time interval $\Delta t$.  The former 
has been calculated in the literature 
but is not directly observable, while the  latter is 
 observable.\\ 
(ii) In Sec.IIID,  $\av{\sigma_{xy}}(t)$ 
has been  displayed together 
with the histograms of the broken bond number $\Delta N_b(t)$ in Fig.2.  
Here stress drops are related to 
increases in the bond breakage. Also 
the plastic strain increment 
$\Delta\gamma_{\rm pl}$  and 
 the elastic strain increment 
 $\Delta\gamma_{\rm el}$ have been phenomenologically 
 introduced in Eqs.(25) and (27).\\ 
(iii) In Sec.IIIE,    the  dislocation gliding 
 and the grain boundary sliding  
 have been studied  in Figs.4 and 5. From the elastic energy 
of a slip  under applied stress in Eq.(30), 
we have argued why the plastic deformations tend 
to be nearly along the $x$ or $y$ axis  
and why the particle displacement vectors are 
clockwise (type C) around  slips along the $x$ axis 
and counterclockwise (type CC) around  slips along the $y$ axis.  
Though these are arguments for slips, 
we have found  the same  preferred directions 
 and  displacements around broken bonds 
 for any $c$.\\
(iv) In Sec.IIIF,  visualization of 
the dynamic and structural 
heterogeneities have been given  
for various $c$. Displayed is 
the bond breakage in the background 
of the disorder variable $D_j$ with   $\Delta t= 
0.2/\gdot$ in Fig.6 and in the background of the orientation angles $\alpha_j$ 
 with $\Delta t= 
 0.5/\gdot$  in Figs.7  and  8.   
The overall  heterogeneity directions are 
 along the $x$ or $y$ axis for any $c$. 
The averaged velocity 
${\bar v}_x(y,t)$ defined by  Eqs.(33) and (34) 
represents the mean  particle motions 
along the $x$ axis. For $\Delta t=0.1/\gdot$ in Fig.9,  
it greatly deviates from  the linear profile 
for any $c$. For $\Delta t=0.8/\gdot$   in Fig.10, 
its deviation still remains for small $c$ but becomes 
 small for not small $c$.\\ 
(v) In Sec.IIIG,  
catastrophic plastic events at  
large stress drops  have  been visualized    
in Figs.11-15 for $c=0.05$ (polycrystal) and  $0.2$ 
(glass) with   $10\le \Delta \le 50$. 
In Fig.14, time-evolution of 
 broken bonds in  sheared glass 
 shows how they appear as short slips 
 and how they aggregate on a time scale of  $\Delta t=10$. 

We give  concluding remarks in the following.\\ 
(1) For any $c$,   plastic  events   
extend  over a mesoscopic area ($\gg \sigma_1$) and  
occur in a short time,  
as visualized  on a time scale of $\Delta t= 50$ 
in  Figs.11 and 12. In crystal and polycrystal 
plasticity is due to 
 slip  formation  and 
grain boundary sliding with various sizes \cite{dislocation,postmortem}. 
In glass, slip lines are short but collectively 
appear  in a narrow region,  as visualized  in Fig.14. 
 Subsequent plastic events 
tend to multiply occur in such  fragile areas 
to form shear bands, which 
can extend thoughout the system for our (still small) system 
size  on much longer  time scales of order $10^3$ 
as in Figs.6-9.   On time scales 
longer than $10^4=1/\gdot$, the correlations of 
plastic events to  earlier ones  
still persist for small $c$ but 
disappear in glass as in Fig.10. 
This  long-range  cooperativity   
propagates  in short times and persist 
for long times decaying slower  in polycrystal  
than in glass.\\
2) Molecular glassy materials  
behave as  elastic bodies on short time scales, 
though mesoscopically  inhomogeneous elastic moduli 
 give rise to irregular deformations
   \cite{Yoshimoto,Barrat-small}. 
   Large-scale intermittent 
   release of the elastic energy  at high stress 
is a common   feature of  plasticity   
in crystal, polycystal, and glass. 
See Figs.5 and 14, where 
nonlinear deformations  
propagate rapidly as long or short slips depending on 
$c$.\\  
(3) We remark on the dynamic heterogeneity 
in glass without shear 
\cite{Takeuchi,Hiwatari,Harrowell,
yo,Kob,Donati,Glo,Biroli,Dol,Kawasaki,Ha,HamaOnuki}. 
Jammed particle configuration changes are first triggered  
in the form of    chains of 
broken bonds \cite{yo,yo1}  or stringlike clusters  
\cite{Kob,Donati,Glo} in microscopic  times, 
which are  mostly around boundaries of 
small remaining crystalline regions  
 \cite{HamaOnuki,Kawasaki}. Subsequently,  
the clusters of the broken bonds  accumulate to form  mesoscopic 
 heterogeneities  on long time scales.  
To confirm this process,  some papers 
have visualized 
 time-evolution of the dynamic heterogeneites  
\cite{yo,yo1,Dol,PTP},  where 
 succesive broken bonds or active regions mostly 
overlap or are adjacent to each other.  
 Applied shear intensifies this tendency, 
eventually leading to organization of 
shear bands, as 
illustrated  in this paper.  
 
(4) In the melting phenomena
in two dimensions, we 
remark that  analogous large-scale  heterogeneities 
 emerge both  in the structural disorder 
and in the dynamics \cite{2Dmelting}.

\begin{acknowledgments}
The authors are grateful to T. Hamanaka, T. Araki, T. Uneyama,  S. Yukawa, 
and A. Furukawa  for valuable discussions.
Some of the numerical 
calculations were carried out on NEC SX8 at
YITP in Kyoto University. 
This work was supported by Grants-in-Aid 
for scientific research 
on Priority Area ``Soft Matter Physics" 
and  the Global COE program 
``The Next Generation of Physics, Spun from Universality and Emergence" 
of Kyoto University 
 from the Ministry of Education, 
Culture, Sports, Science and Technology of Japan. 
H. S. was supported by the Japan Society for Promotion of Science.
\end{acknowledgments}

\vspace{2mm} 
{\bf Appendix: Microscopic Expressions for Average Stress and 
Forces  to Walls}\\
\setcounter{equation}{0}
\renewcommand{\theequation}{A\arabic{equation}}

Our system is composed of the particles 
unbound and bound to the top  and bottom  
walls in the regions $-0.6L<y<-0.5L$ and $0.5L<y<0.6L$. 
   The total potential energy 
of  our system is written as Eq.(5).
 From the  equations of motion  we  derive 
the following equation, 
\be 
\frac{d}{dt}\sum_{j\in {\rm all}} m_j y_j\dot{x}_j 
= \Pi^{0}_{xy} - \sum_{\cal B} 
\sum_{j\in {\cal B}}\bigg 
[y_j \frac{\p u_j}{\p {x}_j}+\zeta_{\cal B}
 m_j y_j (\dot{x}_j-v_{\cal B})\bigg], 
\en 
where  
$\ddot{x}_j= d^2x_j/dt^2$, $\dot{x}_j= d x_j/dt$,   
${\p u_j}/{\p {x}_j}=K [x_j-X_j(t)]$, and $v_{\cal B}=
\pm \gdot L/2$. 
The  subscript 
$\cal B$ denotes the top or bottom boundary. 
We introduce the space integral of the $xy$ component of 
the stress of all the  particles,  
\be 
\Pi^{0}_{xy}= \sum_{j\in {\rm all}}  m_j \dot{x}_j\dot{y}_j
-\sum_{j,k\in {\rm all}} \frac{x_{jk}y_{jk}}{2r_{jk}}\phi'_{jk},
\label{eq:AppStress}
\en 
where  $\dot{y}_j= d y_j/dt$, 
 $r_{jk}=|{\bi r}_j-{\bi r}_k|$, and 
$\phi'_{jk}=\p \phi_{jk}/\p r_{jk}$.

The total forces  acted by the walls 
to the fluid along the $x$ axis are  
given by $F_{\rm top}$ and 
$F_{\rm bot}$ in Eq. (\ref{eq:Ftb}). 
 From Eqs.(A1) and (A2) 
 the average stress 
$\av{\sigma_{xy}}(t)$ in Eq. (\ref{eq:Stress}) 
 and the wall stress $\sigma_{\rm w}(t)$ 
in Eq. (13)  are related   as 
\bea 
&&\hspace{-10mm} 
L^2[ \sigma_{\rm w}(t) 
-\av{\sigma_{xy}}(t)] 
= \frac{d}{dt}\sum_{j\in {\rm all}}  m_j y_j\dot{x}_j -
\Delta \Pi^{0}_{xy} 
\nonumber\\
&& \hspace{-1.2cm}
+ \sum_{\cal B} \sum_{j\in {\cal B}}
(y_j\mp\frac{L}{2}) \frac{\p u_j}{\p x_j}
+  \sum_{\cal B} \sum_{j\in {\cal B}}
 \zeta_{\cal B} m_j y_j (\dot{x}_j-v_{\cal B})  
 . 
\label{eq:AppFluc}
\ena 
where   $-$ is for the top layer  
and $+$ is for the bottom layer in the second line. 
Here we write the contribution to $\Pi^{0}_{xy}$ from the 
bound particles  as     
 \be 
\Delta \Pi^{0}_{xy}=\sum_{\cal B} \bigg[
  \sum_{j\in {\cal B}}  m_j \dot{x}_j\dot{y}_j
-\sum_{{j}~{\rm or}~{k\in {\cal B}}}\frac{x_{jk}y_{jk}}{2r_{jk}}\phi'_{jk}.
\bigg]
\en
In the right hand side of Eq.(A3), 
the first term is the time derivative of  the sum 
$\sum_{j\in {\rm all}}  m_j y_j\dot{x}_j $ of  all the particles, 
while the other terms involve the bound particles. 
Thus the first term is dominant in the limit of large $L$. 
This is indeed the case in our simulation. 
That is,  in the right hand side, the  third   term is  
less than $10\%$ of the first term 
and the second and fourth terms are much smaller than the third  
at any time.


\begin{references}
\bibitem{band} 
In this paper,we do not consider 
fracture or brittle behavior.
We use  the term ``shear band''
widely to represent quasi-linear (in 2D) mesoscopic 
objects  where shear strains 
are localized.
\bibitem{fraction} The minimums of the LJ potentials in Eq.(1) 
are  at $r=2^{1/6}\sigma_{\alpha\beta}$, 
so another nearly equivalent  
definition of the volume fraction is  
$\phi= D (N_1\sigma_1^2+N_2\sigma_2^2)/N$, where  
$D= 2^{1/3}\pi/4 \cong 0.9895$ is close to unity.
\bibitem{temp} To be precise, for $\gdot=10^{-4}$,  
the temperature in the middle 
region became  higher than the boundary value 
$0.2$  by  a few $\%$ \cite{PTP}.  
We also performed simulations with $N=81000$, 
where the temperature inhomogeneity increased considerably 
  at high shear with 
 $\dot{\gamma} \gs  10^{-3}$.
\bibitem{variance} In Ref.\cite{HamanakaShear} 
with $N=10^3$ 
we calculated  the time average of 
$[\av{\sigma_{xy}}(t)- \bar{\sigma}_{xy}]^2 $ 
in plastic flow, denoted by  $\cal V$. 
The  ${\cal V}^{1/2}$ is the typical fluctuation amplitude. 
Comparison of the data in Ref.\cite{HamanakaShear} 
and those in the present simulation 
with $N=9\times 10^3$ (not shown here) 
indicates  ${\cal V}\propto L^{-2}$. 
See Ref.\cite{Pro} also for this point. 
Note that $L^2{\cal V}$ 
is  equal to the  time average  of 
the  quantity  
$L^{-2} 
\int d{\bi r}_1\int d{\bi r}_2
\av{\delta\Pi_{xy}({\bi r}_1,t)\delta\Pi_{xy}({\bi r}_2,t)}$
for the $xy$ component 
 of the 
 stress tensor $\Pi_{xy}({\bi r},t)$\cite{Onukibook}.
\bibitem{climb} Dislocation 
gliding involves motions of a relatively 
small number of particles, 
while the other form of dislocation 
motions, climbing, 
requires motions of many particles 
 and is not easily induced \cite{Friedel}.
\end{references}

\end{document}